\documentclass[prb,amsfonts,amssymb,floats,twocolumn,superscriptaddress,aps]{revtex4-1}
\usepackage[utf8]{inputenc}
\usepackage[T1]{fontenc}
\usepackage{amsmath}
\usepackage{float}
\usepackage{color}
\usepackage{braket}
\usepackage{bm}
\usepackage{dsfont}
\usepackage{adjustbox}
\usepackage{graphicx}
\usepackage{bbm}
\usepackage{tikz}
\usetikzlibrary{patterns}

\newcommand{\red}[1]{#1}

\newcommand{\iu}{\mathrm{i}} 
\newcommand{\eu}{\mathrm{e}} 
\newcommand{\du}{\mathrm{d}} 
\newcommand{\pv}{\mathrm{p.v.}}

\newcommand{\Uone}{\mathrm{U(1)}}

\newcommand{\srio}{Sr$_2$IrO$_4$}
\newcommand{\nio}{NiO}

\newcommand{\heq}{\mathcal{H}_\mathrm{eq}}
\newcommand{\qu}{\mathrm{q}}
\newcommand{\cl}{\mathrm{c}}

\newcommand{\bvec}[1]{{\bm{#1}}}

\newcommand{\uvec}[1]{{\underline{\underline{#1}}}}

\DeclareMathOperator{\diag}{diag}

\graphicspath{{./}{figs/}}

\allowdisplaybreaks

\begin{document}

\title[]
{
Optical excitation of magnons in an easy-plane antiferromagnet:\\Application to \srio{}
}
\author{Urban F.P. Seifert}
\affiliation{Institut f\"ur Theoretische Physik,
Technische Universit\"at Dresden, 01062 Dresden, Germany}
\affiliation{Kavli Institute for Theoretical Physics, University of California, Santa Barbara, CA 93106, USA}
\author{Leon Balents}
\affiliation{Kavli Institute for Theoretical Physics, University of California, Santa Barbara, CA 93106, USA}
\affiliation{Canadian Institut for Advanced Research, Toronto, Ontario, Canada M5G\,1M1}


\date{\today}

\begin{abstract}
We study the interaction of a (classical) light field with the magnetic degrees of freedom in the two-dimensional antiferromagnet \srio. The reduced space group symmetry of the crystal allows for several channels for spin-operator bilinears to couple to the electric field. Integrating out high-energy degrees of freedom in a Keldysh framework, we derive induced effective fields which enter the equations of motion of the low-energy mode of in-plane rotations which couple to the out-of-plane magnetization. Considering a pump-probe protocol, these induced fields excite magnetization oscillations which can subsequently probed, e.g. using Kerr rotation.
We discuss how the induced fields depend on polarization and frequency of the driving light, and our study applies  to both resonant and non-resonant regimes. Crucially, the induced fields depend on the two-magnon density of states, thus allowing for further insight into properties of the magnetic excitation spectrum. Furthermore, these effects rely upon (weak) magnon-interactions, and so are beyond a ``Floquet magnon'' description.
\end{abstract}


\maketitle


\section{Introduction}

Ultrafast optics provide a powerful means to probe and manipulate
electronic materials.  Two well-explored mechanisms of light-matter
interaction in ultrafast optics are through interband transitions
(electron-hole excitations), and through optically active phonon modes.  A much
less explored alternative is to use ultrafast radiation to directly
excite spin excitations. Recent experiments (see below) show that
this can indeed be achieved in Mott insulating antiferromagnets, and
the goal of this paper is to develop a theory for this mechanism of
ultrafast excitation. \cite{orbcurrpol,ultrfa_rmp,dyfeo3}

\red{\subsection{Experimental setup, theory goals and results}}

We consider the following measurement protocol.  A short and intense ``pump''
pulse of radiation is first applied at a frequency comparable to the
magnetic exchange $J$, and within the optical gap.  The dynamics after
the pulse is then probed using a second laser for measurements on
times $t$ much longer than the intrinsic exchange time $t \gg
\hbar/J$.  On this longer time scale, a description solely in terms of
slow modes, whose dynamics is generically semi-classical, is
appropriate.  The creation of magnetic excitations in the form of magnons by
irradiation with short laser pulses has previously been demonstrated in the
antiferromagnet NiO.\cite{fiebig_nio1,fiebig_nio2}  In that work, the
initial pump pulse was treated phenomenologically using free-energy
arguments as an effective magnetic field -- an ``inverse Faraday
effect'' -- while an appropriate
semi-classical picture was applied to the dynamics during the probe
period.

\begin{figure}[!tb]
\includegraphics[width=\columnwidth,clip]{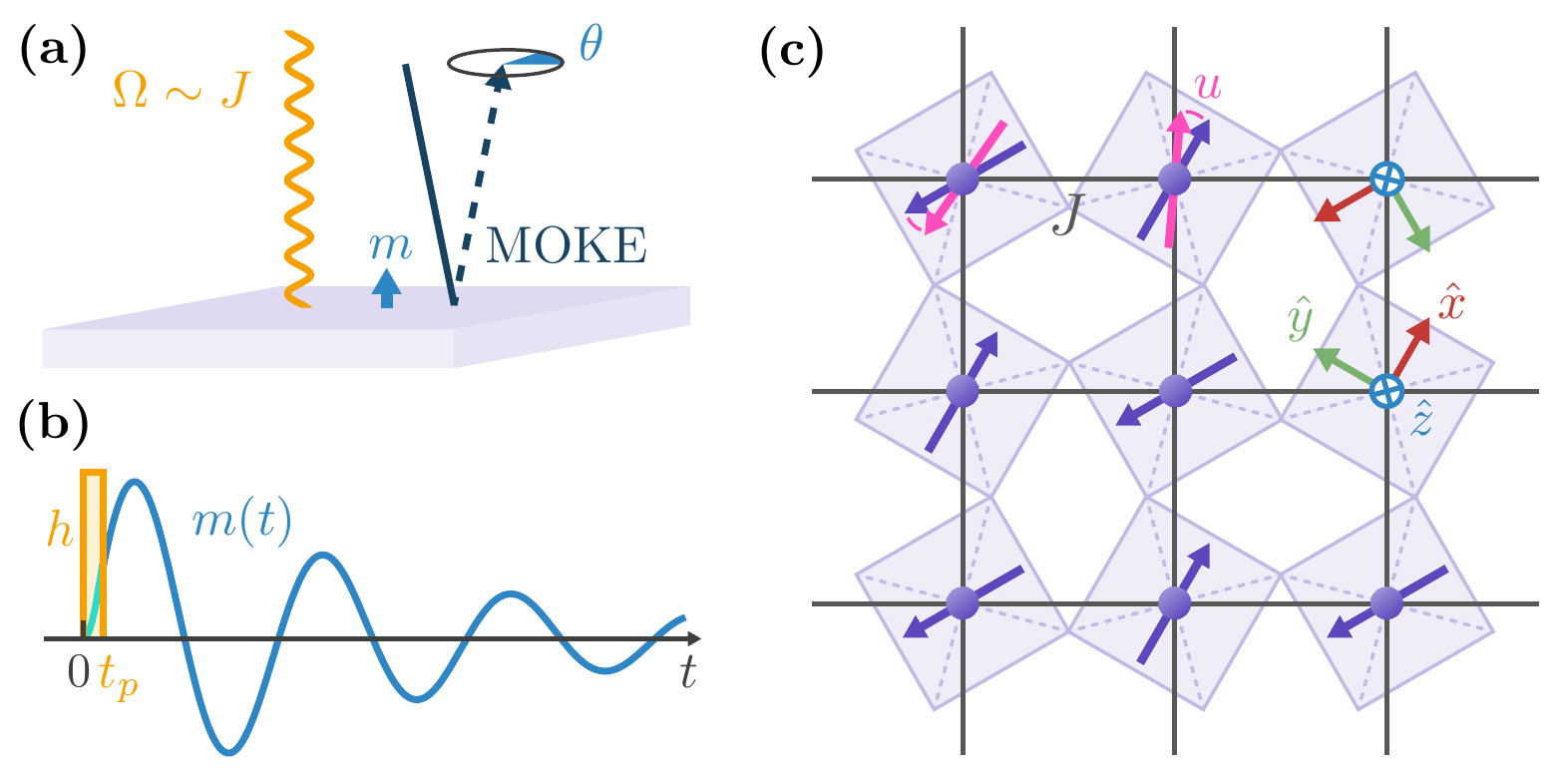}
\caption{
  (a) Measurement protocol consisting of a short pump pulse (yellow line) with frequency $\Omega$ and a probe beam (dark blue line). The polarization rotation $\theta$ of the refracted light is proportional to the magnetization of the sample as a result of the Magneto-optical Kerr effect (MOKE).
  (b) The electric field of the pump induces an effective field $h$ which drives the magnetization out of equilibrium (cyan curve), setting initial conditions for oscillations and relaxation of $m$ according to the equilibrium equations of motion (blue curve).
  (c) Crystal structure of a basal plane of \srio\ with rotated IrO$_6$ octahedra (shaded) forming a square lattice, and $J_\mathrm{eff}=1/2$ moments in equilibrium (global frame), and local coordinate frames used for spin-wave theory. The low-energy in-plane mode corresponds to small fluctuations of the in-plane ordering angle, parametrized by the parameter $u$.
}
\label{fig:introfig}
\end{figure}

Here we {\em derive} the effect of the pump pulse from a microscopic
treatment of the coupling of the electric field of the light to the
spins.  This replaces the free energy arguments, and
gives specific predictions for the magnitude of the effective field of
Refs.~\onlinecite{fiebig_nio1,fiebig_nio2} on polarization, frequency, etc.
Importantly, because the pump pulse is of short duration and frequency
comparable to $J$, the physics during this period is in no way
semi-classical and requires a full quantum treatment.  Prior quantum
theory of the mechanism of an inverse Faraday effect was based on
modeling off-resonant light to a {\em few-level}
system,\cite{pershan65} which is appropriate for isolated atoms but
not collective spin excitations. The present work elevates
this to a full theory of quantum magnons interacting with the pump laser.

We approach this problem in the specific milieu of \srio{}, a
paradigmatic spin-orbit Mott insulator comprising a planar square
lattice antiferromagnet.  \srio{} is attractive for its large magnon
bandwidth, small spin, two dimensionality, and the large spin-orbit
coupling (SOC) which is inherent to the magnetic Ir$^{4+}$ ion.  The
latter is advantageous as SOC is required to connect the spin
polarization and with the orbital polarization of light.  We consider
an easy-plane antiferromagnetic model for the basal planes in \srio{}.
In equilibrium the fluctuations around the Neél ordered state can be
described in linear spin-wave theory in the form of magnon
excitations, the spectrum of which is correctly reproduced in the
low-energy limit by classical equations of motion.

These equations are modified during the duration of the pump by the
optical electric field. \red{The derivation of these modifications from a
microscopic model of the magnetic excitations in the material is the main objective of this paper.} 
For a spatially uniform system, we obtain
\begin{align} \label{eq:EOM_field_gamma-1}
  \partial_t u &= \chi^{-1} m -\frac{u}{\tau_u} -  h_m, \nonumber \\
  \partial_t m &= - \kappa u - \frac{m}{\tau_m} + h_u,
\end{align}
\red{where our principle result is the expression of $h_m,h_u$ in terms of the applied light field.}
These equations of motion describe $\bm{k}=0$ magnons.
Here $u$ is a small fluctuation of the in-plane angle of the spin away from the ordering axis, and $m$ is the out-of-plane magnetization, which  appears because $m$ is the conjugate variable to $u$, and hence generates in-plane rotations.
The parameter $\chi$ is the out-of-plane susceptibility, $\kappa$ is proportional to an in-plane anisotropy, and $\tau_u,\tau_m$ are phenomenological relaxation times that are only important during the ``probe'' period.  
The modifications due to the electric field are the Zeeman-like field $h_m$ and a ``dual'' field $h_u$, both of which are quadratic in the applied field.

\red{ \subsection{Summary of methods and outline}

We briefly summarize the methodology by which we obtain this result, and which is the principle subject of the main text.
A single basal plane of \srio{} contains $J_\mathrm{eff}=1/2$ moments on a square lattice.
The equilibrium Hamiltonian can obtained by symmetry analysis as
 \begin{align} \label{eq:ham}
  \mathcal H_\mathrm{eq} = \sum_{i\in A} \sum_{\mu=\pm x,\pm y} \Big[ &J_{xy} \left(
               {\sf S}_i^x {\sf S}_{i+\mu}^x + {\sf S}_i^y {\sf
               S}_{i+\mu}^y \right) + J_{z} {\sf S}_i^z {\sf S}_{i+\mu}^z  \nonumber \\
               +& D \bm{\hat{z}}\cdot {\sf S}_i \times
               {\sf S}_{i+\mu} 
                  \Big],
\end{align}
where we take $J_{xy} > J_z > 0$ corresponding to an antiferromagnetic Heisenberg coupling with an easy-plane anisotropy, while $D$ denotes the strength of the Dzyaloshinskii-Moriya interaction,\cite{jakha09} and we neglect a small in-plane anisotropy to be discussed later.
Considering the coupling of spins to the perturbing electrical field, we find that time-reversal symmetry mandates the electrical field to couple to a \emph{bilinear} of spin operators, generally written as
\begin{equation}
  \mathcal{H}_\mathrm E = \sum_{\substack{\mu=x,y\\ \alpha,\beta \in \{x,y,z\}}} \sum_{i j} g_{ij}^{\mu \alpha \beta}  E_\mu \mathsf{S}_i^\alpha \mathsf{S}_j^\beta,
\end{equation}
with the tensorial structure of the couplings $g_{ij}^{\mu \alpha \beta} $ made explicit in Eq. \eqref{eq:ham_E}.

The equilibrium spin dynamics (i.e. $E_\mu \equiv 0$) of the system is conveniently investigated by employing a large-$S$ framework to expand about the classical antiferromagnetically ordered state.
To this end, we pick a staggered local reference frame
\begin{equation}
  \label{eq:1}
  \vec{S}_i \rightarrow \epsilon_i \left(S_i^x \hat{n} +  S_i^y
    \hat{z}\times \hat{n} \right) +
  S_i^z \hat{z},
\end{equation}
in which the classical (antiferromagnetic) ground state corresponds to a ferromagnetic configuration with spins aligned along the $x$-axis.
We make use of the Holstein-Primakoff representation to represent the spin operators in terms of bosonic operators. In first non-trivial order of $1/S$ (i.e. linear spin-wave theory), the Hamiltonian is quadratic in the bosons and diagonalization readily yields the spin-wave dispersion
\begin{equation} \label{eq:magnon_disp}
  E_{\bvec{k}} = 2 J S \sqrt{\left( 2- \gamma_{\bvec k} \right) \left(2+(1-\delta) \gamma_{\bvec k}\right)},
\end{equation}
where $\gamma_{\bvec{k}} = \cos k_x + \cos k_y$. 
Similarly, one may expand the interaction $\mathcal{H}_\mathrm E$ in $1/S$, where we include up to three-boson terms.

The strategy to analyze the non-equilibrium interacting boson problem
is to define an appropriate path integral, and integrate out ``fast''
degrees of freedom (large momentum, high frequency) to obtain an
effective description of the low energy dynamics.  Importantly, the high frequency
light fields affect the low-frequency dynamics because modes of
different frequencies are coupled through cubic boson interactions.

In order to obtain the effective low-energy action of the system both
in equilibrium \emph{and} during the pump (i.e. in the presence of a
time-dependent electric field $E_\mu(t)$), we employ a real-time (Keldysh) path integral
\begin{equation}
  \mathcal Z = \int\! \mathcal{D}\left[a_+,a_-\right] \eu^{\iu \mathcal S},
\end{equation}
where $a_+$ and $a_-$ denote bosonic fields on the forward and backward branches of the real-time contour employed in the Keldysh formalism and the action $\mathcal S$ contains contributions both from the non-interacting spin-wave theory Hamiltonian and the perturbation $\mathcal H_\mathrm E$.
Using a cutoff to separate low-energy and high-energy modes and path-integral averaging over the latter then yields a low-energy action $\mathcal S = \int \! \du t \, \du^2 \bm{x} \, S^{-1} \, \mathcal L$ where the Lagrange density $\mathcal L$ is conveniently expressed in terms of the low-energy variables $u$ and $m$ which constitute coherent bosonic modes [see \eqref{eq:low-energy-m-u} for an explicit definition]. 


Since the free (quadratic) action for the Holstein-Primakoff-Bosons does not mix momenta, the equilibrium low-energy Lagrangian $\mathcal{L}_0$ is straightforwardly obtained as the small-momentum part (i.e. $\bm{k} \simeq 0$) of the bare equilibrium action.
We further include a small experimentally observed easy-axis anisotropy $\Gamma > 0$ [see also Eq.~\eqref{eq:gamma_pot}] which violates the in-plane $\Uone$ symmetry of the Hamiltonian \eqref{eq:ham} and thus allows for relaxation of $m$ (which is a conserved quantity of $\mathcal{H}_\mathrm{eq}$).
Including the low-energy Lagrangian due to this in-plane anisotropy $\mathcal{L}_\Gamma$, we find the full equilibrium low-energy Lagrange density 
\begin{equation} \label{eq:l0-leff}
  \mathcal{L}_0 + \mathcal{L}_\Gamma = \left[m_\qu \partial_t u_\cl - u_\qu
                  \partial_t m_\cl \right] - \frac{1}{\chi} m_\cl m_\qu - \kappa u_\cl u_\qu.
\end{equation}
On the contrary, the interaction with the electric field $\mathcal{H}_\mathrm E$ is a perturbation to the bare equilibrium action with three-boson interaction terms which lead to the mixing of momenta.
We thus employ the Keldysh path integral introduced above to integrate out high-momentum modes perturbatively to second order in the electric field, generating an effective low-energy action which describes the coupling of a single slow mode to an effective (induced field) $h$.
We may represent the process of integrating out high-energy modes diagrammatically, with the slow mode (straight line) and external electric fields (dashed lines) appearing as sources, and the internal lines being high-momentum propagators to be averaged over, thus yielding an effective vertex between a slow mode and two electric fields, 
\begin{equation} \label{eq:full-feyn}
  \mathcal{L}_\mathrm{eff} = 
  \begin{tikzpicture}[baseline={([yshift=-.5ex]current bounding box.center)}]
    \path (1,0) coordinate (c);
    \path (c) ++(30:0.3) coordinate (a1); 
    \path (c) ++(30:0.6) coordinate (a2);
    \path (c) ++(-30:0.3) coordinate (b1); 
    \path (c) ++(-30:0.6) coordinate (b2);
    \path (c) ++(180:0.3) coordinate (l1);
    \path (c) ++(180:0.6) coordinate (l2);
    \draw[pattern=north west lines] (c) circle (0.3);
    \draw[dashed] (a1)--(a2);
    \draw[dashed] (b1)--(b2);
    \draw (l1)--(l2);
  \end{tikzpicture}
  =
  \begin{tikzpicture}[baseline={([yshift=-.5ex]current bounding box.center)}]
    \path (1,0) coordinate (c);
    \path (c) ++(0:0.3) coordinate (a1);
    \path (c) ++(0:0.2) coordinate (a0); 
    \path (c) ++(0:0.6) coordinate (a2);
    \path (c) ++(150:0.6) coordinate (b); 
    \path (c) ++(180:0.3) coordinate (l1);
    \path (c) ++(210:0.6) coordinate (l2);
    \draw (a0)--(l1);
    \draw[dashed] (a1)--(a2);
    \draw[dashed] (l1)--(b);
    \draw (l1)--(l2);
  \end{tikzpicture}
  \;\;+\;\;
  \begin{tikzpicture}[baseline={([yshift=-.5ex]current bounding box.center)}]
    \path (1,0) coordinate (c);
    \path (c) ++(0:0.3) coordinate (a1); 
    \path (c) ++(0:0.6) coordinate (a2);
    \path (c) ++(150:0.6) coordinate (b); 
    \path (c) ++(180:0.3) coordinate (l1);
    \path (c) ++(210:0.6) coordinate (l2);
    \draw (c) circle (0.3);
    \draw[dashed] (a1)--(a2);
    \draw[dashed] (l1)--(b);
    \draw (l1)--(l2);
  \end{tikzpicture}
  .
\end{equation}
The effective action due to the presence of electric fields and fluctuations effects can then be put in the form
\begin{align}
  \label{eq:seff}
  \mathcal{L}_{\mathrm{eff}} = h_m m_\qu + h_u u_\qu + \frac{u_\qu u_\cl}{\tau_u} - \frac{m_\qu m_\cl}{\tau_m},
\end{align}
where $h_{m,u}$ are the induced effective fields obtained by explicitly evaluating above diagrams, which is done in Sec.~\ref{subsec:deriv_eff_fields}, and we have also included phenomenological relaxation terms.
Demanding the full action with $\mathcal{L} = \mathcal{L}_0 + \mathcal{L}_\Gamma + \mathcal{L}_\mathrm{eff}$ to be extremal then yields the equations of motion for the classical variables $u_\cl,m_\cl$ as given in \eqref{eq:EOM_field_gamma-1}.

We note that above considerations show that the finite induced fields necessarily require {\em interactions} between magnons, because the electric field itself couples to large momentum modes whose energy is comparable to the pump frequency, and only anharmonic terms provide the mode-coupling that generates an effect upon the zero momentum fields.
Thus these results cannot be obtained from any ``Floquet magnon'' description. 
} 

It should be emphasized that our approach is inherently quantum, and allows to study both resonant and off-resonant contributions to the induced effective fields for linear and circularly polarized light.
In particular, connections between spectral properties of the magnons and the induced fields entering the low-energy dynamics can be elucidated. In short, our calculations give several definite predictions for future experiments \red{which are discussed in Sec.~\ref{sec:expapp}}, and a methodology that can be much more widely applied.

The remainder of the paper is organized as follows.
In Sec.~\ref{sec:model}, we give a full exposition of the model for \srio{}, including the coupling between spins and electric fields.
Then, in Sec.~\ref{sec:light-induced-low}, we apply the spin wave expansion to obtain an effective bosonic Hamiltonian, formulate the non-equilibrium dynamics in a Keldysh path integral, and carry out a perturbative diagrammatic derivation of the low energy equation of motion.
We conclude in Sec.~\ref{sec:summary-outlook} with a a discussion of observable consequences, and an outlook for further applications and extensions. Several appendices give technical details of calculations summarized in the main text.
 
\section{Model and Symmetries} \label{sec:model}

For concreteness, we consider a single plane of \srio{}, in which
corner-sharing IrO$_6$ octahedra form a square lattice of Ir$^{4+}$
ions, as depicted in Fig.~\ref{fig:introfig}.
As is well established, spin-orbit coupling leads to the emergence of
$J_\mathrm{eff} = 1/2$ moments with effective nearest-neighbor
Heisenberg interactions.\cite{jakha09,bjk1,bjk2} 

\subsection{Symmetries} \label{sec:symms}

Symmetry places an important constraint upon the physics.
As the corner-sharing octahedra in \srio{} are rotated at an angle $\theta \simeq 13^\circ$, the unit cell on the square lattice of the
$J_\mathrm{eff}=1/2$ is doubled. A full account of the three dimensional structure places \srio{} in the I$4_1/$acd space group.\cite{crawf}
However, because of the strong two-dimensionality of \srio{}, well-confirmed by experiment, it is appropriate to consider the full set of symmetries of a single basal IrO$_2$ plane. These are generated by \red{(i)} a fourfold rotation $C_4^z$ \red{about the $z$-axis perpendicular to the plane}, \red{(ii)} inversion $\mathcal{I}$, \red{(iii)} a horizontal mirror operation $\sigma_h$ and \red{(iv)} a screw operation along the $x$ axis, \red{consisting of a lattice translation $x \to x+1$ paired with a $\pi$-spin rotation about the $x$ axis)}. Not all such operations are true symmetries of the three-dimensional structure, but they are good approximate symmetries, and a model analysis of the exchange couplings based on these symmetries is highly successful in explaining the equilibrium properties of \srio{}, so we proceed with the same symmetry assumptions here.

\subsection{Hamiltonian}

Many prior studies have established the minimal spin Hamiltonian $\mathcal H_\mathrm{eq}$ for \srio{} as given in \eqref{eq:ham}, where we have neglected the influence of a small in-plane anisotropy $\Gamma$ on the bare Hamiltonian.
The spin Hamiltonian can also be obtained from microscopic considerations taking into account the electronic structure of Sr$_2$IrO$_4$ and perturbation theory around the Mott limit.\cite{khaliu12}   

For further analysis, it will prove convenient to work with spin operators $\vec{S}_i$ in a local frame in which the spins order collinearly. To this end, we perform a staggered rotation ${\sf S}_i^\pm = \eu^{\pm \iu \epsilon_i \theta} S_i^\pm$, ${\sf S}_i^z = S_i^z$, where $\epsilon_i = \pm 1$ for $i \in A$ ($B$) sublattices.\cite{jakha09}
The Hamiltonian thus becomes 
\begin{align} \label{eq:heq}
  \mathcal \heq = J \sum_{i \in A} \sum_{\mu=\pm x,\pm y} \bigg[\frac{1}{2} \left(S^+_i S^-_{i+\mu} + S^-_i S^+_{i+\mu} \right) \nonumber\\ + (1-\delta) S^z_i S^z_{i+\mu} \bigg],
\end{align}
where $J = \sqrt{J_{xy}^2 + D^2}$ and $\theta$ is determined by $\tan 2\theta = - D/J_{xy}$, and we have introduced the dimensionless anisotropy parameter $\delta = 1- J_z / J $.
\red{It is easily seen that in this local reference frame, the spin Hamiltonian given in \eqref{eq:heq} corresponds to an XXZ model with a $\Uone$ symmetry of picking the in-plane ordering axis.}

\subsection{Coupling to electric field}

We assume that the dominant coupling of the light to the spins is
through the electric field.  This is reasonable for a strongly
spin-orbit coupled system, owing to the weakness of the magnetic
component of the laser field.  
Since $\vec E$ is even under time reversal, the electric field needs to couple to a bilinear of spin operators.
The most general such interaction Hamiltonian linear in $\vec{E}$
and involving nearest-neighbor spins that is invariant under the symmetries \red{(i)-(iv) given in Sec.~\ref{sec:symms}} is
\begin{align} \label{eq:ham_E}
  \mathcal H_\mathrm{E} & = \sum_i  \Big[ g_1 \epsilon_i \left[ E_x \left({\sf S}^y_i
        {\sf S}^x_{i+x} + {\sf S}^x_i {\sf S}^y_{i+x}\right) - (x\leftrightarrow y) 
        \right]
        \nonumber \\
  &+ g_2 \epsilon_i \left( E_y {\sf S}_{i}^x {\sf S}_{i+x}^x - E_x {\sf
  S}_i^y {\sf S}_{i+y}^y\right) \nonumber \\
  & + g_3  \epsilon_i \left( E_y {\sf S}_{i}^y {\sf S}_{i+x}^y - E_x {\sf
    S}_i^x {\sf S}_{i+y}^x\right) \nonumber \\
  & + g_4 \epsilon_i {\sf S}_i^z \left( E_y {\sf S}_{i+x}^z - E_x {\sf
    S}_{i+y}^z\right) \nonumber \\
    & + g_5 \left( E_y \, \bm{\hat{z}}\cdot {\sf S}_i \times {\sf
      S}_{i+x} - E_x \,\bm{\hat{z}}\cdot {\sf S}_i \times {\sf
      S}_{i+y}\right) \Big].
\end{align}
It should be emphasized that a doubling of the unit cell (due to the
staggered rotation of the octahedra) is crucial to obtain finite $g_1,
\dots, g_4$. 

The microscopic origin of the couplings in \eqref{eq:ham_E} can be elucidated by noting that the Hamiltonian defines a spin-dependent electric polarization $\vec P$ through $\vec P = -\partial \mathcal{H}_\mathrm E / \partial \vec E$.
The spin-dependent electric polarization in transition metal oxides
can be obtained in a microscopic calculation by considering the
electronic structure of a TM-O-TM dimer and expressing the matrix
elements of $\vec P$ in the Mott limit through spin
operators,\cite{knb} and has recently been applied to the relevant
geometry of a bond angle $\alpha \neq \pi$, giving rise to the
additional terms introduced above (see also Sec.~\ref{sec:expapp} and Appendix~\ref{sec:MEcoup}).\cite{bolens}


\section{Light-induced low-energy dynamics}
\label{sec:light-induced-low}

\subsection{Transformation to bosons}
\label{sec:transf-bosons}

We concentrate on low temperature,
for which it is appropriate to expand small fluctuations around the magnetically ordered
ground state. We use the standard  Holstein-Primakoff  $1/S$ expansion ($S$
is the spin magnitude), which transforms the spin problem order by
order into one of bosons, whose normal modes in the quadratic
approximation are magnons, or spin waves.  We start with the
equilibrium problem with $\vec{E}=0$.  We parametrize the classical ground state as
$\vec S_i = \epsilon_i S \hat{n}$, where $\hat{n}=(\cos\phi,\sin\phi,0)$ is a unit vector in
the x-y plane.  The angle $\phi$ parametrizes a
$\Uone$ freedom in picking the Neél ordering axis of \eqref{eq:heq}.
This freedom is ultimately split by in-plane anisotropy, but we will
simply leave $\phi$ as a free parameter for the present analysis.  

As usual, we choose a ferromagnetic local frame for spin-wave theory by transforming to new spin variables \red{as defined in Eq.~\eqref{eq:1}} so that the classical ordered state simply corresponds after the transformation to ferromagnetic alignment along the $\hat{x}$ axis.
Then we make the Holstein-Primakoff transformation which expresses the spin operators in terms of bosonic operators $a$ as $S^+_i = S^z_i - \iu S^y_i = \sqrt{2 S - n_i} a_i$, $S^-_i = (S^+_i)^\dagger = a_i^\dagger \sqrt{2 S - n_i}$ and $S^x_i = S - n_i$,
where $n_i = a_i^\dagger a_i$. The resulting Hamiltonian $\heq$ can be
expanded as $\heq = \heq^{(0)} + \heq^{(2)} + \mathcal{O}(1/S^0)$,
where $\heq^{(n)}$ contains $n$ bosonic operators.  While $\heq^{(0)}$
corresponds to the classical ground-state energy, the quadratic part
$\heq^{(2)}$ can be diagonalized by means of a Bogoliubov
transformation $a^{\vphantom\dagger}_{\bm k} = \cosh \vartheta_{\bm k}
c_{\bm k}^{\vphantom\dagger} + \sinh\vartheta_{\bm k} c_{- {\bm
    k}}^\dagger$ (see Eq.~\eqref{eq:2}).  This yields
\begin{equation}
  \label{eq:3}
\heq^{(2)} = \sum_{\bvec{k}} E_{\bvec{k}} c_{\bvec{k}}^\dagger
c^{\vphantom\dagger}_{\bvec{k}} + \mathrm{const},
\end{equation}
with the spin-wave dispersion $E_{\bvec k}$ given in \eqref{eq:magnon_disp}.
It is convenient to combine the bosonic operators in a Nambu spinor $\psi_{\bvec{k},\sigma} = (a_\bvec k, a_{-\bvec{k}}^\dagger)^T$, with $[\psi_{\sigma,\bvec{k}},\psi_{\sigma',\bvec{k}'}] = \epsilon_{\sigma,\sigma'} \delta_{\bvec{k},-\bvec{k}'}$. Analogous to $\heq$, we also expand the light-spin interaction Hamiltonian $\mathcal{H}_\mathrm E$ in the Holstein-Primakoff bosons. We finally obtain a schematic expansion of the form
\begin{align} \label{eq:he-schem}
  \mathcal{H}_\mathrm E = \sum_\mu E_\mu (t) \big[\Phi^{1,\mu}_{A} \psi_A + \Phi^{2,\mu}_{A,B} \psi_A \psi_B  \nonumber \\ + \Phi^{3,\mu}_{A,B,C} \psi_A \psi_B \psi_C + \mathcal{O}\left(1/S^0\right) \big],
\end{align}
\red{where the $\Phi^{m,\mu}_{A,\dots}$ are vertex functions which in general depend on momentum, the couplings $g_1,\dots,g_5$ and parameters which specify the classical ground state (i.e. the angles $\phi$ and $\theta$), and we use composite indices $A=(\bvec k, \alpha)$ etc. (with appropriate summations implicit) for brevity.
For a more detailed expression of $\mathcal H_\mathrm{E}$, detailing the tensorial structure, we refer the reader to Eq.~\eqref{eq:he-full} in Appendix \ref{sec:deriv}.}
Importantly, we have included the leading cubic interaction between
magnons induced by the electric field. As mentioned in the
introduction, coupling between modes of different momentum, which
occurs only through such interactions, is necessary for the electric
field to induce effects on the long wavelength modes.  We remark that
an additional source of magnon interactions arises from intrinsic
exchange interactions already present in equilibrium.  However, we
find that these give subdominant contributions to the field-induced
terms in the equations of motion for the slow modes, and hence ignore
them here.  

Finally, note that the unit cell doubling restricts the momenta in the original
unit cell to be conserved \red{only} up to the (magnetic ordering) wave vector
$\bvec{Q}=(\pi,\pi)$.

\subsection{Keldysh formulation}

The presence of the time-dependent electric field in \eqref{eq:ham_E}
takes the system out of equilibrium.  We formulate the problem using
Keldysh path-integral representation.  To this end, the bosonic fields
are time-evolved along a folded contour from $t=-\infty$ to $t=\infty$
and back, corresponding to time-evolving the density matrix of the
system. Denoting bosonic fields on the forward (backward) contour by
$a_\pm$, the Keldysh path integral of the system is given by
$\mathcal{Z} = \int \mathcal{D}[a_+,a_- ] \exp(\iu \mathcal S)$ with
the action
\begin{equation}
  \mathcal S = \sum_{s=\pm} s  \int\! \du\mathrm{t}\, \left\{ \sum_i
    \bar{a}_{s,i} \iu\partial_t a_{s,i}  - \mathcal H [\{\bar{a}_{s,i},a_{s,i}\}]\right\},
\end{equation}
where the Hamiltonian $\mathcal H = \heq^{(2)} + \mathcal{H}_\mathrm{E}(t)$ is given by the quadratic spin-wave Hamiltonian and the time-dependent interaction Hamiltonian \red{[given in Eqs.~\eqref{eq:3} and \eqref{eq:he-schem}, respectively]}.
In practice, it is useful to perform a Keldysh rotation to classical and quantum fields $a_\mathrm{c,q} = (a_+ \pm a_-)/\sqrt{2}$, where the classical field can acquire a non-zero expectation value.
The classical equations of motion for $a_\cl$ then emerge as solutions to the saddle-point equations
\begin{equation} \label{eq:cl_saddle}
  \frac{\delta \mathcal{S}}{\delta a_\mathrm{q}} = 0, \quad a_\mathrm{q} = 0.
\end{equation}

\subsection{Slow modes and low energy equation of motion}
\label{sec:slow-modes}

The analysis of slow modes proceeds by considering slowly varying
components of bose fields in the Keldysh path integral.  The procedure
is to work in the Fourier basis of fields in momentum space, and
integrate out those fields with momentum above some cut-off.
Formally, the resulting effective action governs the dynamics of the
remaining small momentum modes.  In the Fourier representation, terms
up to quadratic order in boson operators do not mix momenta.
Therefore, within linear spin wave theory, the effective action is
simply equivalent to the bare action for the small momentum modes.  It
already in equilibrium contains those terms which govern the
non-dissipative dynamics of the magnons.

\red{To find an appropriate low energy effective action, it is convenient to
express the boson creation and annihilation operators in terms of more
physical combinations that represent the collective variables of the
easy-plane antiferromagnet.  Specifically, since the gap of the
out-of-plane mode can be made sufficiently large by increasing the
anisotropy parameter $\delta$, we consider the latter to be a {\em
  high} energy mode, and focus just on the easy-plane degrees of
freedom.  When spatial variations of the ordered state are small and slowly
varying, we may write the classical spin configuration as
$\vec S_i = (S \epsilon_i , \epsilon_i u(\bm{x}_i), m(\bm{x}_i))^T$ in terms of a phase
$u$ and angular momentum $m$, which correspond to the phase of the
in-plane ordering axis and out-of-plane magnetization,
respectively.  We assume $u(\bm{x})$ and $m(\bm{x})$ vary slowly on
the lattice scale.  Then, comparing with the Holstein-Primakoff
expansion, we identify the magnetization $m(\bm{x}_i)$ and the in-plane phase $u(\bm{x}_i)$ 
\begin{equation} \label{eq:low-energy-m-u}
  m(\bm{x}_i) = \sqrt{\frac{S}{2}}(a_i^{\vphantom\dagger}  + a_i^\dagger), \quad u(\bm{x}_i) = \iu \sqrt{\frac{S}{2}} \left(a_i^{\vphantom\dagger} - a_i^\dagger \right).
\end{equation}
The bosonic commutation relations imply that $u$ is the canonically
conjugate variable to $m$, as also obtained from the classical Poisson
bracket $\{ S^y, S^z \} = S^x$ with $\vec S$ given above. 
We may rewrite the small momentum Keldysh action obtained from keeping
up to quadratic terms in the bosons as a function of $u$ and
$m$ and their gradients.  For our purposes, we consider spatially
uniform fields and responses, and it is sufficient for the low energy
effective action to work to $0$th order in a spatial gradient
expansion, i.e. to neglect any terms with spatial derivatives of $m$
or $u$.
The resulting effective action is then of the form $\mathcal{S} = \int\! \du t\,\du^2\bvec{x}\, S^{-1}\mathcal{L}$, where we make the $S$-scaling explicit.
For the equilibrium XXZ model \eqref{eq:heq}, the Lagrange density $\mathcal L$ is given by
\begin{equation}
  \mathcal{L}_0 = \left[m_\qu \partial_t u_\cl - u_\qu
                  \partial_t m_\cl \right] - \frac{1}{\chi_0} m_\cl m_\qu,
\end{equation}
with the inverse susceptibility $1/\chi_0 = 2(2-\delta) JS $.
Note that at the present stage, $m$ is a constant of motion corresponding to the U(1) symmetry of the model under in-plane rotations.

We expect two types of corrections to this ``bare'' action, arising in
the process of integrating out higher modes due to the
cubic and higher boson interactions.  First, already in equilibrium,
in the absence of any applied electric fields, there are
``self-energy'' corrections which renormalize the magnon spectrum and,
more importantly, which give rise to a finite magnon lifetime. In a
general model, moreover, microscopic interactions
violating the $\Uone$  symmetry will allow relaxation of the uniform
magnetization. In the model at hand, we hence include the $\Gamma$ interaction
\begin{equation} \label{eq:gamma_pot}
  \mathcal{H}_\Gamma = \Gamma \sum_{\langle i j \rangle} S^x_i S^x_j - S^y_i S^y_j,
\end{equation}
giving rise to an in-plane easy-axis anisotropy.
Recent studies have identified a pseudo-spin Jahn-Teller effect as the origin of the experimentally observed anisotropy (see also Sec.~\ref{sec:expapp}).\cite{khaliu19} Note that $\Gamma \ll J$, so that the renormalization of the magnon spectrum is small at higher energies and can be neglected in \eqref{eq:magnon_disp} for our purposes.
Second, when electric fields are applied, the bare
action will also be corrected by generation terms which drive an
initially zero magnetization to become non-zero, and otherwise
deform the equilibrium steady state.  
Given these considerations, the full low-energy Lagrangian can be separated as
\begin{equation}
  \mathcal L = \mathcal{L}_0 + \mathcal{L}_\Gamma + \mathcal{L}_\mathrm{eff},
\end{equation}
where the bare Lagrange density for the low-energy modes obtained in the harmonic approximation upon the inclusion of $\mathcal{H}_\Gamma$ is given in \eqref{eq:l0-leff}. 
Clearly, the violation of the $\Uone$ symmetry leads to dynamics of the magnetization, where $1/\chi =4(2-\delta) J S + 4 \Gamma S$ defines a moment of inertia (given by a renormalized inverse susceptibility) and $\kappa = 8 \Gamma S$ is seen to correspond to a torsion constant.
Fluctuation effects generated by integrating out higher energy modes then lead to the effective Lagrangian $\mathcal{L}_\mathrm{eff}$ given in \eqref{eq:seff},
where $h_m$ and $h_u$ are the effective fields we seek to compute here, and $\tau_u$ and $\tau_m$ are relaxation times.


Applying the saddle point conditions $\partial{\mathcal{L}}/\partial
m_q = \partial{\mathcal{L}}/\partial u_q = 0$ and taking $m_q=u_q=0$,
we obtain Eq.~\eqref{eq:EOM_field_gamma-1} of the introduction.  
These equations of motions govern the dynamics of the $\bvec{k}=0$ mode in equilibrium at $T=0$. Note that for the case of $T>0$, stochastic noise terms would need to be added to \eqref{eq:EOM_field_gamma-1} for the fluctuation-dissipation theorem to hold. However as we are interested in $T\to 0$ (for which we expect the spin-wave approach to be reliable), we will neglect these terms henceforth. 

}
Computing the eigenvalues of the system \eqref{eq:EOM_field_gamma-1} for the case of $h_m = h_u= 0$ yields
\begin{equation}
  \lambda_\pm = -(\gamma_m + \gamma_u) \pm \iu\sqrt{\omega_0^2 - (\gamma_m - \gamma_u)^2}
\end{equation}
with $\omega_0^2 = \kappa / \chi$ and $2 \gamma_{u,m} = 1/ \tau_{u,m}$. It thus becomes clear that a finite $\tau_u$ causes a faster relaxation, however counteracts the frequency renormalization due to damping through a finite $\tau_m$. The limit $\tau_u \to \infty$ recovers conventional Brownian motion for $m$. For our purposes, with damping expected to be small, it is sufficient to follow a more heuristic approach by keeping only the damping in the equation of motion for $m$ and work with a generalized relaxation time $\bar\tau^{-1} = \tau_u^{-1} + \tau_m^{-1}$ and, if necessary, afterwards renormalize the eigenfrequency $\bar\omega = \mathrm{Im}[\lambda_+]$ appropriately.

Considering the full dynamics of \eqref{eq:EOM_field_gamma-1}, the effective fields can be seen to act as driving forces which take the magnetization $m$ out of its initial position of rest, $m=\partial_t m = 0$.
While the pump of duration $t_p$ is active, $m$ then evolves according to the EOM \eqref{eq:EOM_field_gamma-1} in the presence of the effective fields. After the pump is switched off, $m$ evolves according to the equations of motion of the free (damped) harmonic oscillator [i.e. $h_{u,m} = 0$ in \eqref{eq:EOM_field_gamma-1}]. However, by continuity, the initial conditions for the harmonic time-evolution of the magnetization have to be obtained by integrating the equations of motion during the pump.
For a detailed discussion, we refer to Appendix \ref{sec:intpump}. In the ultrafast regime in which the pump is short compared to the period of the oscillations and the relaxation time $t_p \ll \gamma^{-1}, \omega_0^{-1}$, we find that, after the pump, the initial conditions for the free magnetization oscillations are given by
\begin{align}
  m(t_p^+) &\simeq \bar{h}_u t_p \label{eq:mhu} \\
  \partial_t m (t_p^+) &\simeq \kappa \bar{h}_m t_p, \label{eq:dtmhm}
\end{align}
where $\bar{h}_{u,m}$ are the (constant) pulse strengths and $\kappa = 16 \Gamma S$.
In the ultrafast regime, the effective field $h_m$ therefore acts as an impulse which provides an initial velocity, while $h_u$ provides an initial amplitude of $m$. We discuss experimental consequences in Sec. \ref{sec:expapp}.


\subsection{Derivation of effective fields} \label{subsec:deriv_eff_fields}

We now calculate the effective fields in Eq.~\eqref{eq:seff} by including  the time-dependent perturbation \eqref{eq:ham_E} and integrating out the fast modes of the system. 
\red{To this end, we consider the light-spin interaction Hamiltonian \eqref{eq:he-schem} in the Keldysh formalism, perform a Keldysh rotation and Bogoliubov transform to obtain (again employing a schematic notation in the interest of clarity)
\begin{align}
  \mathcal{S}_\mathrm{E} = - \int\! \du t \sum_\mu &E_\mu(t) \left\{ 2 \hat{\Phi}^{2,\mu}_{A,B} \hat{\psi}_A^\mathrm{q} \hat{\psi}_B^\mathrm{q} \right. \nonumber\\ &+\left.  2^{-1/2} \hat{\Phi}^{3,\mu}_{A,B,C} \left(\hat{\psi}_A^\mathrm{q} \hat{\psi}_B^\mathrm{q} \hat{\psi}_C^\mathrm{q} + 3 \hat{\psi}_A^\mathrm{q} \hat{\psi}_B^\mathrm{c} \hat{\psi}_C^\mathrm{c} \right) \right\}, \label{eq:se-schem}
\end{align}
where $\hat{\psi} = \Lambda^{-1} \psi = (c_{\bvec k}, c_{-\bvec k}^\dagger)^T$ denotes the Bogoliubov-rotated Nambu spinor for which the equilibrium magnon action $\mathcal{H}_\mathrm{eq}^{(2)}$ is diagonal (for details, we again refer to Appendix \ref{sec:deriv}).
}
We define fast modes $\hat{\psi}$ as those with an energy larger than a cut-off $\omega_c$, i.e. those with $\omega > \omega_c$.  The low energy modes appear as sources, as far as the integration over the fast fields is concerned.
In addition, the electric fields themselves are also sources, since they are externally imposed classical variables, although they are not ``slow''.
\red{We can then treat $\mathcal{S}_\mathrm{E}$ perturbatively with respect to the free magnon action which follows from $\mathcal{H}_\mathrm{eq}^{(2)}$. Integrating out high-energy modes can be defined via a diagrammatic expansion, in which both slow fields and electric fields appear as external legs [solid and dashed external legs in Eq. \eqref{eq:5}, respectively].} 
\red{The effective fields $h_m$ and $h_u$ are linearly related to $h_\alpha$ [with $\alpha=1,2$, see also Eq.~\eqref{eq:h_um_alpha}]  which needs to couple to a single quantum bosonic mode $\psi_\alpha^{\mathrm q}$ to enter the classical equations of motion.
One then finds that the terms corresponding to the effective fields $h_m$ and $h_u$ have one external solid line and two external dashed lines at the lowest quadratic order in the electric fields.
The two Feynman diagrams which lead to induced fields (quadratic in $E_\mu$) for the low-energy $u,m$ modes are shown in Eq.~\eqref{eq:full-feyn}.
We find that the xxxxxx-level diagram is trivial, since energy and momentum conservation require the internal line to be a slow mode and thus fall below the cutoff, and thus the effective action is given by a single loop integral,
\begin{equation}
  \mathcal{L}_\mathrm{eff} = S h_\alpha \psi_\alpha^\mathrm{q}  \label{eq:5}
  = \;
  \begin{tikzpicture}[baseline={([yshift=-.5ex]current bounding box.center)}]
    \path (1,0) coordinate (c);
    \path (c) ++(0:0.3) coordinate (a1); 
    \path (c) ++(0:0.6) coordinate (a2);
    \path (c) ++(150:0.6) coordinate (b); 
    \path (c) ++(180:0.3) coordinate (l1);
    \path (c) ++(210:0.6) coordinate (l2);
    \draw (c) circle (0.3);
    \draw[dashed] (a1)--(a2);
    \draw[dashed] (l1)--(b);
    \draw (l1)--(l2);
  \end{tikzpicture}\;\;
  .
\end{equation}}
Note that the absence of terms that are cubic in the bosons in the equilibrium model \eqref{eq:heq} (this a result of a $\mathbb{Z}_2$ symmetry $S_i^y \rightarrow - S_i^y, S_i^z \rightarrow -S_i^z$  in the XXZ model) and therefore the lack of cubic vertices places a strong constraint on possible diagrams. Cubic boson terms could be induced in the action by including an in-plane anisotropy term, which however is expected to be small, so that any additional resulting diagrams can be expected to be parametrically small, as well.


\red{For the loop integral in \eqref{eq:5}, we note that the external leg being quantum and the form of the vertices in $\mathcal{S}_\mathrm{E}$ fix a unique assignment of classical and quantum fields. Since $\mathcal{H}^{(2)}_\mathrm{eq}$ is diagonal for the Bogoliubov-rotated fields, the internal propagators for $\hat{\psi}$ are given in terms of free Keldysh (harmonic oscillator) Greens functions of the $c_{\bvec k}$ bosons.
After some algebra (for details see Appendix \ref{sec:deriv})}, one finds
\begin{widetext}
\begin{align}
  \label{eq:h_full}
  h_\alpha = \frac{1}{4}\int \! \frac{\du^2\bm{k}}{(2\pi)^2}\,
  \Bigg[&\mathcal{E}_\mu \mathcal{E}_\nu^* \sum_{\beta,\beta'= \pm 1} \beta' \frac{\hat\Phi^{2,Q,\nu}_{\beta,\beta'}(0,-\bvec{k})
  \tilde\Phi^{3,Q,\mu}_{\alpha,-\beta,-\beta'}(0,\bvec{k},\bvec{Q}-\bvec{k})} { \Omega  + \beta
  E_{\bm{k}} +\beta' E_{\bm{Q}-\bm{k}} + \iu \eta } + \left(\mathcal{E}_\nu
  \leftrightarrow \mathcal{E}_\mu, \Omega \rightarrow
    -\Omega\right)\Bigg]\left(2n_B(\epsilon_{\bm{k}}) + 1\right),
\end{align}
\end{widetext}
where we have rewritten the electric fields in terms of a complex amplitude $\mathcal{E}_\mu$ as $E_\mu(t) = \mathcal{R}[\mathcal{E}_\mu \eu^{\iu \Omega t}]$, and $\eta > 0$ is a small convergence factor.
The effective field $h_\alpha$ is related to the source terms in \eqref{eq:seff} as
\begin{equation} \label{eq:h_um_alpha}
  h_u = \iu\sqrt{2 S} \left(h_- - h_+\right),\quad h_m = \sqrt{2 S} \left(h_+ + h_-\right).
\end{equation}
Using that $\Phi^2 \sim S$, $\Phi^3 \sim \sqrt{S}$ and $E_\bvec{k} \sim S$, we hence find that the induced fields scale as $\mathcal{O}(S^1)$.

We evaluate $h_\alpha$ by focussing on $\beta = \beta' = -1$ in the first term and $\beta = \beta' = +1$ in the second, which corresponds to processes in which a magnon pair is created.
Using the Dirac identity $(x + \iu \eta)^{-1} = - \iu \pi \delta(x) + \pv \, x^{-1}$, each field splits into a energy-conserving (resonant) contribution, proportional to $\int\!\du^2\bvec k \,\delta(\Omega - \epsilon_\bvec{k} - \epsilon_{\bvec{Q}-\bvec{k}})$, and a energy non-conserving contribution which includes the principal value integral $\pv \int\!\du^2\bvec k$, corresponding to virtual transitions.

\begin{figure}[!tb]
\includegraphics[width=\columnwidth,clip]{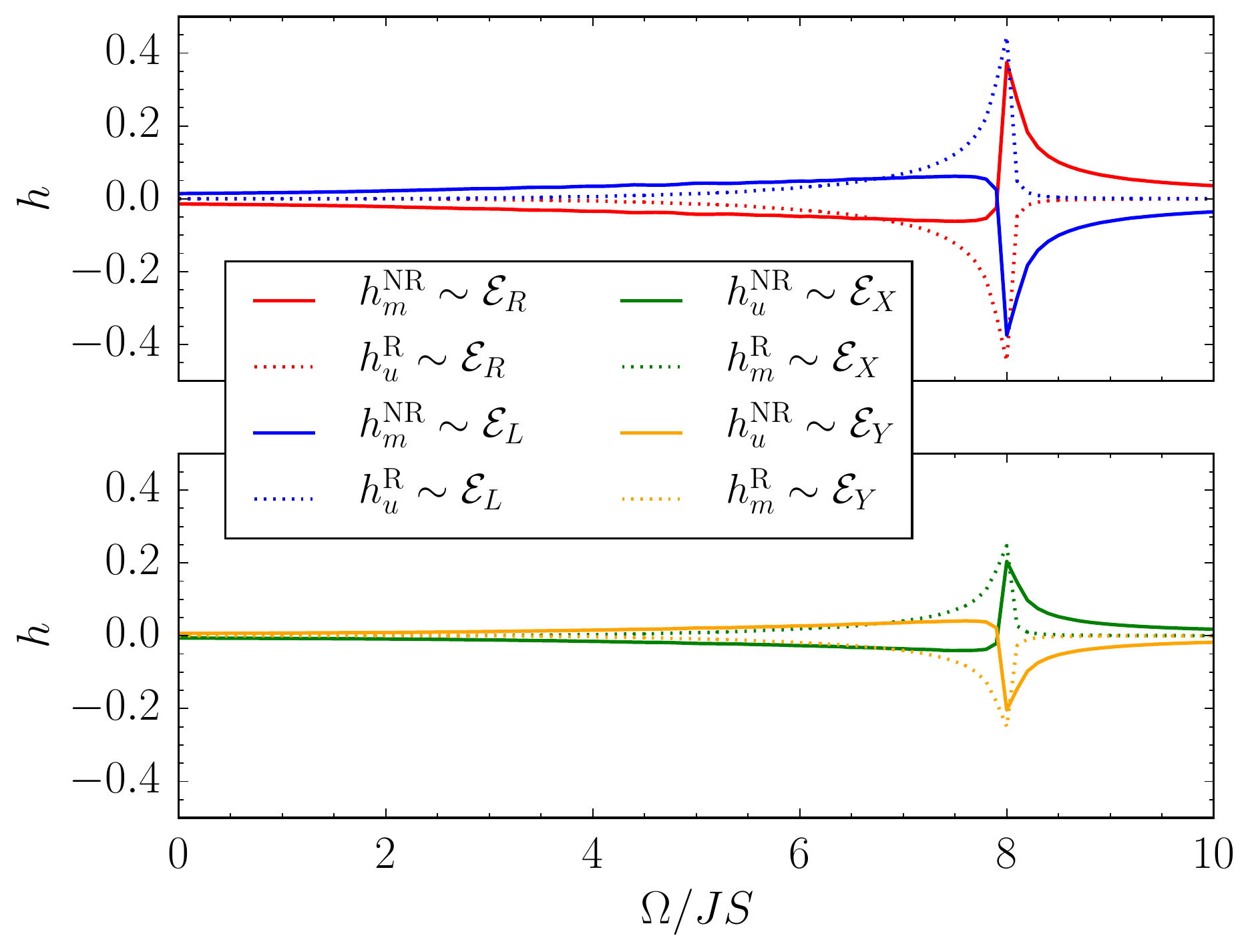}
\caption{
  Effective fields $h_{u,m} = h_{u,m}^\mathrm{R} + h_{u,m}^\mathrm{NR}$ split into resonant (R) and non-resonant (NR) contributions for circular polarizations $\mathcal{E}_{R,L}$ and linear polarization $\mathcal{E}_{X,Y}$,\cite{fn:nonzero,fn:numerics} for couplings $g_i = 1$ except $g_3=-1$.
  Reversing the polarization direction results in field reversal, such that in particlar circularly polarized laser light acts as an effective out-of plane magnetic field corresponding to the inverse Faraday effect.
  The contributions are strongly peaked at $\Omega = 8 JS$ due to the singularity in the two-magnon density of states.
}
\label{fig:omegaScan}
\end{figure}

\begin{figure}[!tb]
\includegraphics[width=\columnwidth,clip]{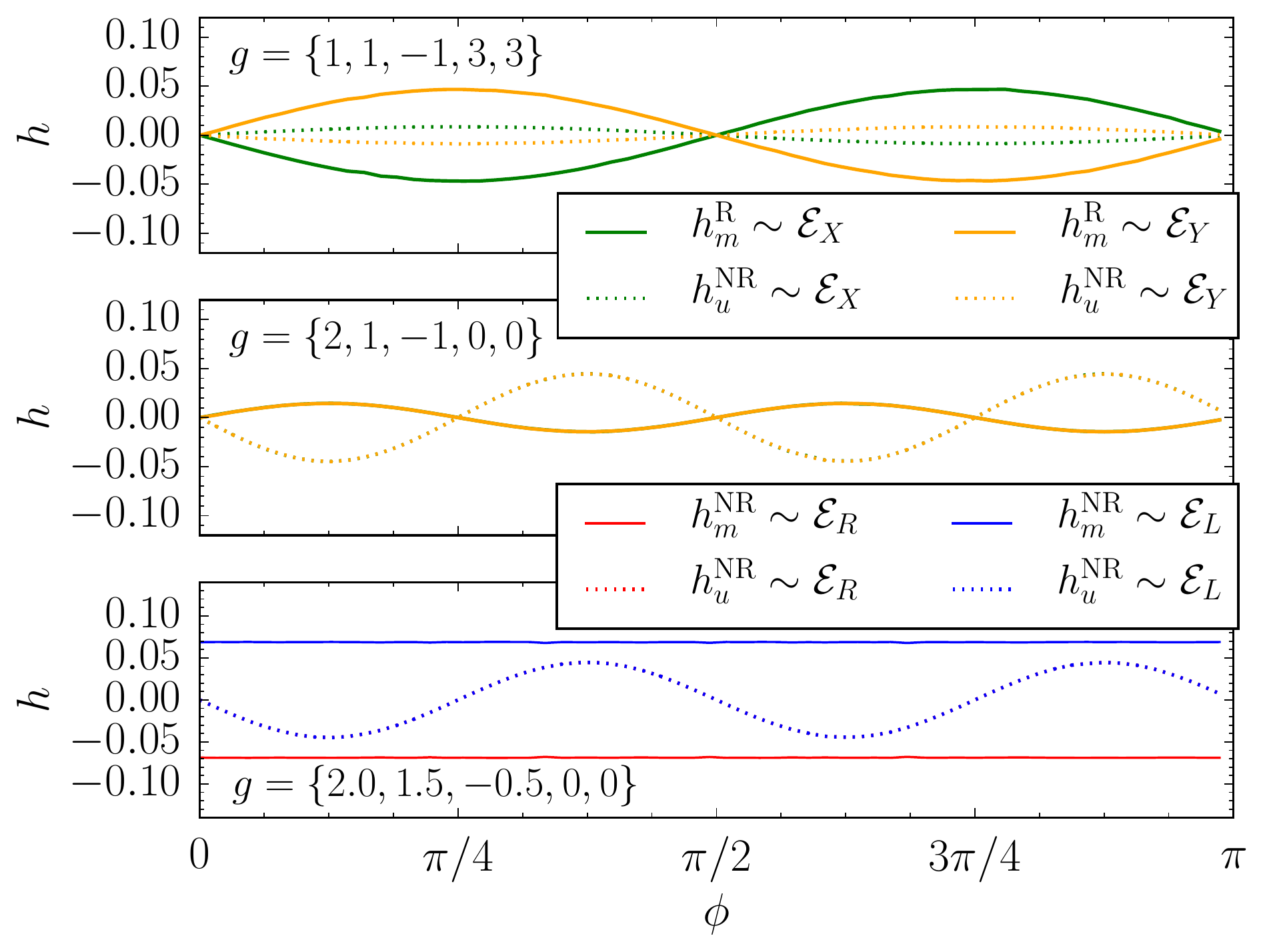}
\caption{
  Dependence of induced effective fields $h_{u,m} = h_{u,m}^\mathrm{R} + h_{u,m}^\mathrm{NR}$ for a linearly polarized pump beam on angle $\phi$ between polarization and magnetic ordering axes for three different sets of magnetoelectric couplings $g$ at $\Omega = 4 J S$.\cite{fn:nonzero,fn:numerics}
  As apparent from Eqs. \eqref{eq:hu_analy} and \eqref{eq:hm_analy}, different contributions to the induced fields can be activated by different choices of $g_i$ and distinguished by their $\phi$-dependence. For a discussion of the different coupling sets used in the panels we refer to the main text.
}
\label{fig:phiScan}
\end{figure}

In order to elucidate the field-dependence of $h_u$ and $h_m$, we expand the vertex functions and dispersion in \eqref{eq:h_full} to first order in the anisotropy $\delta$ and consider long-wavelengths by working to linear order in $\bvec{k}$. The momentum integration then extends over momenta with $\lambda \leq |\bvec k | < \Lambda$, where $\lambda$ is an IR cutoff equivalent to an energy cutoff $\omega_c$ to separate fast from slow modes, and $\Lambda$ is a UV cutoff determined by the microscopic lattice spacing. Upon performing the momentum space integrals, we find it safe to take the cutoff $\lambda \to 0$, yielding
\begin{align}
  h_u &= \mathcal{N}_{u,\Lambda}^{\mathrm{lin}} (\mathfrak{g},J,\Omega,\delta,2\phi,2\theta; \mathcal{E}_x \bar{\mathcal{E}}_x - \mathcal{E}_y \bar{\mathcal{E}}_y; \mathcal{E}_x \bar{\mathcal{E}}_y + \mathcal{E}_y \bar{\mathcal{E}}_x) \nonumber \\
  &+ (4 g_1^2 -(g_2-g_3)^2) \sin 4 \phi \, \mathcal{N}_{u,\Lambda}^\mathrm{int}(J,\Omega,\delta) \, \mathcal{E} \cdot \bar{\mathcal{E}} \nonumber  \\
  &+ g_1 (g_2 - g_3) \mathcal{R}_u (J,\Omega,\delta) \ \iu \mathcal{E} \times \bar{\mathcal{E}}, \label{eq:hu_analy} \\
  h_m &= \mathcal{R}_m^\mathrm{lin} (\mathfrak{g},J,\Omega,\delta,2\phi,2\theta; \mathcal{E}_x \bar{\mathcal{E}}_x - \mathcal{E}_y \bar{\mathcal{E}}_y, \mathcal{E}_x \bar{\mathcal{E}}_y + \mathcal{E}_y \bar{\mathcal{E}}_x) \nonumber \\
  &+ (4 g_1^2 -(g_2-g_3)^2) \sin 4 \phi \, \mathcal{R}^\mathrm{int} (J,\Omega,\delta) \, \mathcal{E} \cdot \bar{\mathcal{E}} \nonumber \\
  &+g_1 (g_2 - g_3) \, \mathcal{N}_{m,\Lambda} (J,\Omega,\delta) \ \iu \mathcal{E} \times \bar{\mathcal{E}} \label{eq:hm_analy},%
\end{align}%
where we use the shorthand $\mathfrak{g}=\{g_i g_j\}\setminus\{g_4^2,g_4 g_5, g_5^2\}$. The functions $\mathcal{N}$ contain numerical prefactors originating from the principal value integral, while $\mathcal{R}$ denotes resonant terms. Explicit expressions for the fields are given in Appendix \ref{sec:analy_fields}.

\subsection{Analysis of effective fields}

The field bilinears $\mathcal{E}_x \bar{\mathcal{E}}_x - \mathcal{E}_y \bar{\mathcal{E}}_y$ and $\mathcal{E}_x \bar{\mathcal{E}}_y + \mathcal{E}_y \bar{\mathcal{E}}_x$ are only finite for linearly polarized light, while the chiral intensity $\iu \mathcal{E} \times \bar{\mathcal{E}}$ is only nonzero for circularly polarized light. 
We note that both fields also contain a term which is proportional to the \red{total} intensity $\mathcal{E} \cdot \bar{\mathcal E}$.
The fields $h_u$ and $h_m$ can thus be induced by either linearly polarized light or circularly polarized light.
It is instructive to change to a circularly polarized basis with $\mathcal{E}_{R/L} = (\mathcal{E}_x \pm \iu \mathcal{E}_y)/\sqrt{2}$, such that the chiral intensity reads
\begin{equation}
  \iu \mathcal{E} \times \bar{\mathcal{E}} = \mathcal{E}_R \bar{\mathcal{E}}_R - \mathcal{E}_L \bar{\mathcal{E}}_L.
\end{equation}
It thus becomes clear that the sign of resonant and off-resonant contributions to the effective fields $h_u$ and $h_m$ respectively (which are proportional to the chiral intensity) can be reversed by inverting the pump helicity, corresponding to the inverse Faraday effect, as previously been obtained in the study of a quantum-mechanical few-level system,\cite{pershan65} and also obtained from macroscopic free-energy based approaches.\cite{pershan67,gaiva08,fiebig_nio2}
Note that the exact antisymmetry under helicity switching is spoiled by the $\mathcal{E}\cdot \bar{\mathcal{E}}$ terms -- however these are in the respective opposite regime (resonant/non-resonant) than the $\iu \mathcal{E} \times \bar{\mathcal{E}}$ and thus only of limited relevance.

We remark that these chiral terms in Eqs.~\eqref{eq:hu_analy} and \eqref{eq:hm_analy} are independent of both the angles $\phi$ and $\theta$ and only nonzero if $g_1 \neq 0$ and $g_2 \neq g_3$. As discussed in Section~\ref{sec:model}, these terms are symmetry allowed due to the doubled unit cell which, in the case of \srio{}, is caused by the staggered rotation of the octahedra.

Since the couplings are likely to be of a similar physical origin, it is reasonable to assume that $g_1,g_2,g_3$ are on a similiar order of magnitude (see Sec.~\ref{sec:expapp} for a discussion of the couplings in \srio{}).
We note that the inverse Faraday effect, i.e. the dependence of the fields on the chiral intensity, is maximized for $g_1 = g_2 = -g_3$. In this limit, the prefactor of $\mathcal{E}\cdot \bar{\mathcal{E}}$ in the effective fields vanishes.
Focussing (for simplicity) on this particularly symmetric case (and also taking $g_4 = g_5=0$) in \eqref{eq:ham_E}, we further also notice that $\mathcal{H}_\mathrm{E}$ in the local basis is independent on $\theta$.
The dependence on $\phi$ in this case can be eliminated by rotating the electric field by $-2 \phi$. This amounts to rotating the electric field polarization with respect to a fixed spin-ordering axis.
Since the chiral intensity is independent under planar rotations of $\vec{\mathcal{E}}$, the corresponding terms in $h_u$ and $h_m$ must not depend on $\phi$.
However we stress that in general varying $\phi$ (i.e. the orientation of the ordering axis) is \emph{not} equivalent to rotating the electric field polarization vector.

A perhaps surprising feature of our results is that the induced
  fields depend not only on the chiral intensity $\iu \mathcal{E}
  \times \bar{\mathcal{E}}$ , but in fact on all possible electric
  field bilinears. In a previous  study of a few-level
  system in Ref.~\onlinecite{pershan65}, only the chiral intensity was shown to
  drive a finite magnetization in the off-resonant limit (while the
  other contributions give rise to additional splittings).  The
  distinction arises because we consider the dynamics within a
  magnetically ordered state, rather than a paramagnet, and
  consequently the magnetization $m$ is coupled to another mode
  ($u$), and driving either $m$ or $u$ can induce a magnetization.
  Consequently we see that linearly polarized light can also induce magnetization.

The non-resonant contributions $\mathcal{N}$ depend on the UV cutoff, which regularizes a logarithmic divergence. We hence numerically evaluate $h_u$ and $h_m$ on a lattice for different sets of couplings $\{g_i\}$.\cite{fn:numerics}
The induced fields as a function of $\Omega$ (at fixed $\phi,\theta$) are shown in Fig.~\ref{fig:omegaScan}, and as a function of $\phi$ for a fixed driving frequency of $\Omega = 4 JS$ in Fig.~\ref{fig:phiScan}.

\begin{figure}[!tb]
\includegraphics[width=.9\columnwidth,clip]{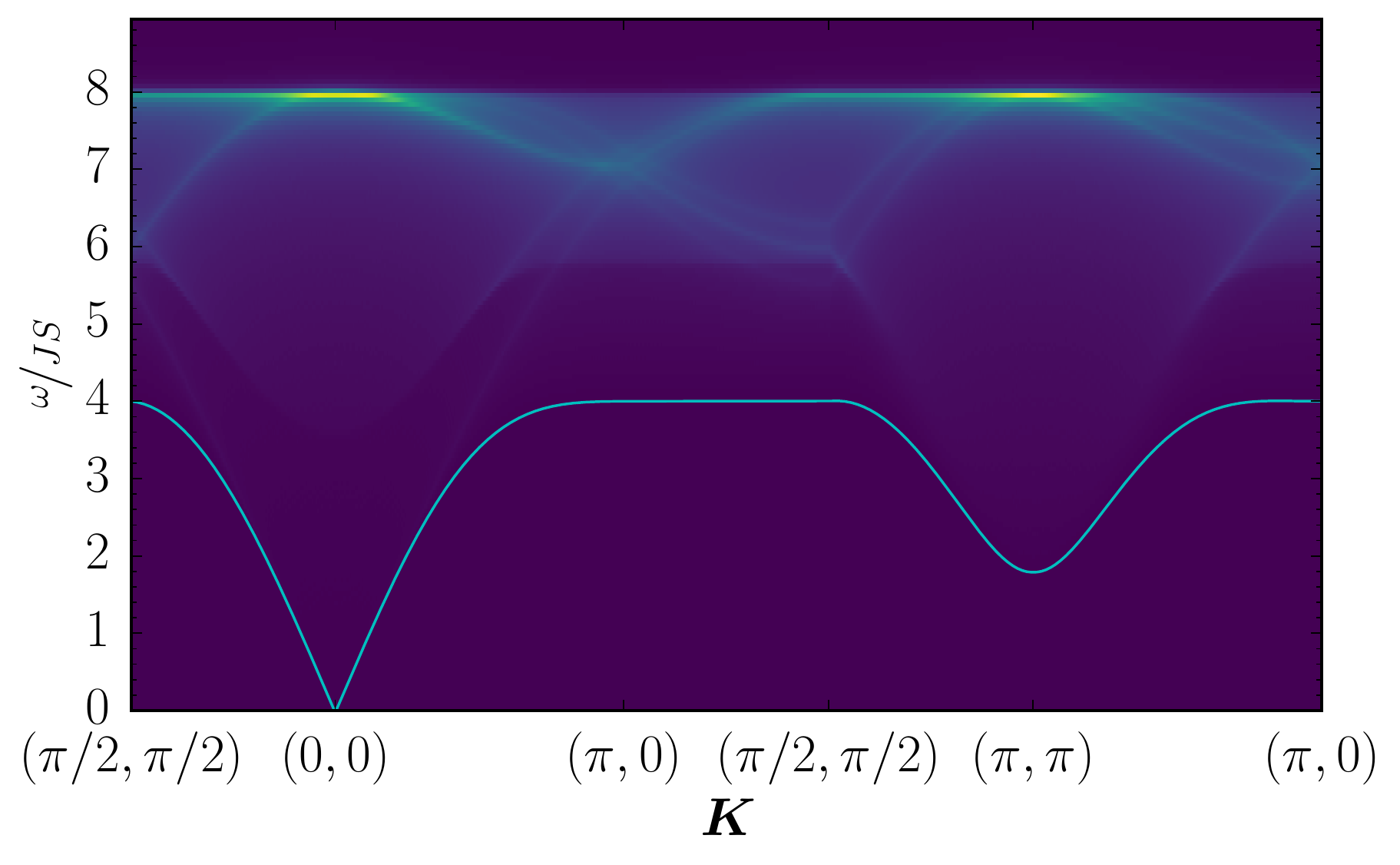}
\caption{
Two-magnon density of states $D_2(\omega,\bvec K)$ as a function of net momentum $\bvec K$ evaluated for $\delta = 0.1$.\cite{fn:numerics} The magnon band maxima at $\omega = 4 J S$ are nested, leading to peaks at $\bvec K = (0,0)$ and $\bvec K = \bvec Q$. Note that $(0,0)$ and $\bvec Q$ become degenerate as $\delta \to 0$.
}
\label{fig:twoMagDOS}
\end{figure}

Notably, we find that the resonant contribution to the fields are strongly peaked and the non-resonant terms undergo a sign change at $\Omega = 8 J S$.
Inspecting the two-magnon density of states (2DOS) $D_2(\omega,\bvec K)$ as a function of energy and net momentum $\bvec K$ in Fig.~\ref{fig:twoMagDOS}, we find that the density of states is singular at $\bvec K = 0$ and at $\bvec K = \bvec Q$ which is the relevant case for the process considered in this study.
This divergence in the 2DOS can be attributed to the fact that the maxima of the dispersion with $\mathrm{max} \, E_\bvec k = 4 J S$ are nested with wavevector $\bvec Q$.

While the components of the induced effective fields proportional
  to the chiral intensity $\iu \mathcal{E} \times \mathcal{E}^\ast$
  are independent of $\phi$, contributions which are only
  non-vanishing for linearly polarized light oscillate harmonically
  with argument $2 \phi$. Terms which are proportional to the
  total intensity $\mathcal{E} \cdot \bar{\mathcal{E}}$ depend on the
  fourth harmonic of $\phi$ (i.e. oscillate with argument $4 \phi$).  As shown in Fig.~\ref{fig:phiScan}, different choices
  of couplings (which we take as free parameters for now) \red{lead to the activation of} the various contributions to the induced fields.  We
  briefly discuss the choices of coupling sets in the figure: (i) The
  set $\{1,1,-1,3,3\}$ maximizes the inverse Faraday effect and
  cancels any contributions proportional the total intensity, and
  induces a $2 \phi$ dependence for linear polarizations since
  $g_4,g_5 \neq 0$.  (ii) The fields due to linearly polarized light
  for the set $\{2,1,-1,0,0\}$ show a $4 \phi$ dependence and no
  dependence on polarization rotation, since only the total intensity
  proportional to $4 g_1^2 - (g_2-g_3)^2 \neq 0$ in
  Eqs.~\eqref{eq:hu_analy} and \eqref{eq:hm_analy} gives a
  contribution.  (iii) If circularly polarized light is used and
  couplings $\{2.0,1.5,-0.5,0,0\}$ are chosen (which do not correspond
  to the symmetric case discussed above), contributions proportional
  to the chiral intensity are independent of $\phi$, while the
  $\mathcal{E} \cdot \bar{\mathcal{E}}$ terms give rise to a $4 \phi$
  dependence which is invariant under pump helicity switching, akin to
  case (ii). 

\section{Summary and Outlook}
\label{sec:summary-outlook}

We conclude with a discussion of the direct application to experiments, and an outlook for future work.

\subsection{Experimental applications}\label{sec:expapp}

\red{We first summarize the key implications for experiments on \srio{}.
\begin{enumerate}
  \item We explain for out-of-plane magnetization oscillations in this easy-plane magnet, and relate the energetics of the corresponding mode to microscopic parameters.
  \item We show how this mode is excited by the $\vec E$-field of the light, through anharmonic magnon interactions, and provide explicit formulas for the excitation.
  \item We predict strong and specific dependence on the excitation polarization and frequency, on the scale of the two-magnon energy.
  \item We inform our many-body modelling with microscopic calculations of the magneto-electric coupling, providing estimates for all phenomenological parameters in the former.
\end{enumerate}
These results are consistent with existing experiments, and provide definite predictions to be tested in future ones.}
\red{To elaborate, we note that in \srio{} the $J_\mathrm{eff}=1/2$ moments interact via a nearest-neighbor} antiferromagnetic Heisenberg coupling of $J_1 \simeq 60 \,\mathrm{meV}$ (we neglect second-and third nearest neighbor interactions for simplicity) and $\delta \simeq 0.05$.\cite{swdamping} 
Since this is an effective spin-1/2 system there are some non-negligible quantum renormalizations, but we still expect the spin wave approach to be qualitatively correct and a good first quantitative approximation.
In addition, lattice deformations give rise to an in-plane axial anisotropy for the $J_\mathrm{eff}=1/2$ moments via a pseudo Jahn-Teller-effect,\cite{khaliu19,bjk3} which fixes the in-plane Neél ordering axis at an angle $\phi \simeq \pi/4$ to the bonds of the effective square lattice of the Ir atoms.
Choosing an appropriate local frame, the anisotropy potential is given by Eq.~\eqref{eq:gamma_pot}
where a recent estimate gives $\Gamma \simeq 3 \, \mu \mathrm{eV}$.\cite{khaliu19}
Recall that \eqref{eq:heq} is invariant under rotations of the Neél ordering axis.
Focussing on the equilibrium dynamics of \eqref{eq:EOM_field_gamma-1}, corresponding to $h_m = h_u=0$, i.e. the light switched off, a finite $\Gamma$ couples the dynamics of $u$ and $m$, therefore lifting the in-plane magnon gap at $\bvec k =0$, as the phase $u$ oscillates with frequency $\omega_0 = \sqrt{\kappa / \chi} \approx 8 S \sqrt{\Gamma J}$.\cite{khaliu19}
With above values, we find that $\omega_0 \simeq 2 \mathrm{meV}$, well below the gap for the out-of-plane mode at $\Delta \simeq 40 \mathrm{meV}$.\cite{bjk3}

\red{As derived explicitly in Sec~\ref{sec:slow-modes}, an ultrafast laser pulse will thus create oscillations of the out-of-plane magnetization in \srio with frequency $\omega_0$ by inducing effective fields which drive the system out of equilibrium and set the initial conditions for dynamics (and relaxation) which is governed by the equilibrium spin model.}

In the following discussion, we focus on \red{ultrafast} pulses, so that $h_u$ provides an initial amplitude, while $h_m$ acts as an impulse that gives an initial velocity of the oscillations.
In the non-resonant regime, our results in Eqs.~\eqref{eq:hu_analy} and \eqref{eq:hm_analy} show that the field $h_m$ is induced by circularly polarized light, while $h_u$ is proportional to the linearly polarized components.
Importantly, we do not expect an explicit dependence of the effective fields $h_u$ or $h_m$ on the total intensity $\mathcal{E} \cdot \bar{\mathcal{E}}$ in \srio{}, since $\sin 4 \phi \simeq 0$ vanishes for the ordering axis angle $\phi \simeq \pi/4$ (cf. above).
As a result of Eqs.~\eqref{eq:mhu} and \eqref{eq:dtmhm} different excitation mechanisms can be distinguished in time-resolved magnetization $m(t)$ measurements of the sample, e.g. through the magneto-optical Kerr effect,\cite{ultrfa_rmp} showing sine-like oscillations (by giving an initial velocity through $h_m$) or cosine-like oscillations (through an initial amplitude provided by $h_u$).
We note that these results are consistent with the phenomenological treatment by Satoh \textit{et al.} in the context of NiO.\cite{fiebig_nio2}


Crucially, the strength of the induced effective fields depends on the pump frequency through the two-magnon density of states, with pronounced features near the singularity.
Probing various pump frequencies and the subsequent response thus might give further insight into magnon dynamics in \srio{}.

We suggest that our study in comparison with experimental data might also be beneficial in clarifying magnetoelectric couplings in \srio{}, which could also further enhance understanding of the giant magnetoelectric effect observed.\cite{gme}

We have evaluated the magnetoelectric couplings $g_1 \dots g_5$ from the microscopic considerations by Bolens in Ref.~\onlinecite{bolens} using exact diagonalization, obtaining
\begin{equation} \label{eq:mecoup}
  g = e \, a_0 \times \{-8.0 \times 10^{-4}, 0.071,0.053,0.097,-0.089\}
\end{equation}
where $a_0$ is the Bohr radius and $e$ the electric charge (for further details, we refer to Appendix~\ref{sec:MEcoup}).
We emphasize that these results depend sensitively on the orbital polarization integrals, for which the hydrogenlike orbitals employed in this study are only a crude approximation (especially for the spread-out $5d$ orbitals).
Importantly, we note that an anisotropy $g_2 \neq g_3$ (required for a finite inverse Faraday effect) is caused by Hund's coupling and spin-orbit coupling (cf. Appendix \ref{sec:MEcoup}).
In order to compare the terms in Eq.~\eqref{eq:ham_E} with the energy scale of intrinsic exchange interactions, we estimate the electric field strength of $|E| \simeq 30 \, \mathrm{mV} / a_0$,\cite{fn:efieldstrength} so that the typical energy scale for the electric-field spin-spin interaction appears to be on the order of $1 \,\mathrm{meV}$.


 Going beyond the immediate application to \srio{}, we note that magnetization oscillations following an ultrashort laser pulse have been observed in other antiferromagnets with significant magnetic anisotropies and strong spin-orbit coupling, such as DyFe0$_3$ and \nio{}.\cite{dyfeo3,fiebig_nio1,fiebig_nio2}
\nio{} is a fcc antiferromagnet with nearest-neighbor Heisenberg couplings between $S=1$ moments,\cite{hutchings_nio} with a single-ion anisotropy of $D_1 = 0.1 \mathrm{meV}$ leading to easy-plane behavior, and thus bears similarities to \srio. A small additional in-plane single-ion anisotropy $D_2 = 0.005 \mathrm{meV}$ further fixes the ordering direction.
Pumping with circularly polarized light, the polarization rotation of the probe beam oscillates with a frequency of $140 \mathrm{GHz}$, corresponding to the excitation of the in-plane mode. Note that main results of our study, such as the phase change of $\pi$ upon switching the pump helicity, and the phase of the oscillations depending on the excitation with linearly or circularly polarized light (corresponding to the different fields), are consistent with the results found in Ref.~\onlinecite{fiebig_nio2}.

\subsection{Outlook}
\label{sec:outlook}

In this work, we have shown that the electrical field of a laser can couple directly to the spin degrees of freedom in the Mott-insulating antiferromagnet \srio. Integrating out high-energy magnons in a non-equilibrium Keldysh framework, we find that this pumping radiation induces effective fields in the classical equations of motion for the low-energy (in-plane) mode, which couples to the out-of-plane magnetization.
\red{Our microscopic treatment allows us to make several key predictions, such as polarization dependence and spectral properties, to be tested in future experiments.}
Our study applies to several other antiferromagnetic systems, for which the optical creation of magnons has been previosuly demonstrated experimentally.\cite{dyfeo3,fiebig_nio1,fiebig_nio2}

For future theoretical studies and experimental applications, we mention that while the induced fields can be formally evaluated also at finite temperatures $T>0$, the underlying Holstein-Primakoff expansion becomes unjustified due to Mermin-Wagner divergences. A reliable account of the temperature dependence of the magnetization oscillations to be observed would therefore likely need to resort to a generalized spin-wave/Schwinger-Boson approach for modelling the magnetic excitations at finite temperatures.\cite{sbmft}  Apart from demonstrating the excitation of magnons in \srio\ and verifying characteristic features of the excitation mechanism, it would be a strong test of the theory to experimentally map out the dependence of the induced fields on the spectral properties of the magnons by varying the frequency of the pump laser.  

The present theory applies to a simple collinear two-sublattice antiferromagnet.  There are several natural extensions.  One obvious one would be to systems with more complex order parameters, such as the planar 120$^\circ$ order of a triangular lattice antiferromagnet.  In such cases, there can be strong cubic anharmonicities in the spin wave Hamiltonian already neglecting magnetic anisotropies and coupling to the light, and based on the calculations in the present paper this may be expected to enhance the effect of a magnon pump.  Another important direction is to consider the effect of optical excitation of topologically non-trivial magnons \cite{topomag1,topomag2} which may possess Berry curvature singularities or non-zero Chern numbers.  This can lead to the presence of edge modes and a mechanism for thermal Hall effects, but the former become depopulated at low temperature since the magnons are excited states, and the latter are exponentially suppressed in Arrhenius fashion for the same reason.  Exciting the topologically non-trivial magnon states directly might lead to enhanced effects.  Considerable recent progress with the electronic analog, i.e. using ultrafast optical methods to probe non-trivial topological properties of electronic bands beyond a linear response regime,\cite{nagaosa16} may serve as inspiration.   The framework presented here provides a firm foundation for future studies in these directions.

\acknowledgements

We thank R. Averitt, A. Bolens, D. Hsieh and S. Zhang for useful discussions, and A. Bolens for the provision of a numerical routine for evaluating magnetoelectric couplings. UFPS was supported by the German Research Foundation (DFG) through the Collaborative Research Center SFB 1143. LB was supported by the Army Research Office MURI grant ARO W911NF-16-1-0361, Floquet engineering and metastable states.  

\appendix

\section{Derivation of the effective fields} \label{sec:deriv}

For reference, we give the full \emph{equilibrium} Keldysh action including Heisenberg exchange $J$, Ising anisotropy $\delta$ and easy-axis anisotropy $\Gamma$ as well as induced fields as source terms,
\begin{align}
  \mathcal S = &\int\! \du t\,\du^2\bvec{x} \, S^{-1} \Big\{ \left[m_\qu \partial_t u_\cl - u_\qu \partial_t m_\cl \right] \nonumber\\
  &- \big[4 (2-\delta) J S + 4 \Gamma S \big] m_\cl m_\qu - 8 \Gamma S u_\cl u_\qu \nonumber\\&+ h_m m_q + h_u u_q\Big\}.
\end{align}
 
\red{\subsection{Perturbation theory for light-spin interaction}}
Expanding the light-spin interaction Hamiltonian in the Holstein-Primakoff bosons $a$ to order $\sqrt{S}$ in the spin size $S$, we find
\begin{widetext}
\begin{align} \label{eq:he-full}
  \mathcal{H}_\mathrm E = \sum_{\bvec A = 0,\bvec Q}\sum_\mu E_\mu (t) \sum_{\bvec k_1,\alpha_1} \Big[ \Phi^{1,\bvec{A},\mu}_{\alpha_1}(\bvec k_1) \psi_{\alpha_1,\bvec k_1}  &+ \Phi^{2,\bvec{A},\mu}_{\alpha_1,\alpha_2}(\bvec k_1, \bvec A - \bvec k_1) \psi_{\alpha_1,\bvec k_1} \psi_{\alpha_2,\bvec{A}-\bvec k_1} \nonumber \\& + \sum_{\substack{\bvec{k}_2,\alpha_2 \\ \bvec{k}_3,\alpha_3}}  \delta_{\sum_i \bvec{k}_i, \bvec{A}}  \Phi^{3,\bvec{A},\mu}_{\alpha_1,\alpha_2,\alpha_3}(\bvec k_1, \bvec k_2, \bvec k_3) \psi_{\alpha_1,\bvec{k}_1},\psi_{\alpha_2,\bvec{k}_2},\psi_{\alpha_3,\bvec{k}_3}\Big],
\end{align}
\end{widetext}
where we have introduced the spinor notation $\psi_{\alpha,\bvec k} = (a_\bvec k, a_{-\bvec k}^\dagger)^T$, and the vertex functions scale as
$\Phi^1 \sim \sqrt{S}^3$, $\Phi^2 \sim S$ and $\Phi^3 \sim \sqrt{S}$.
Note that the staggered rotation leads to a doubling of the unit cell in real space, so that momenta are in principle only conserved up to $\bvec Q = (\pi,\pi)^T$.
\red{The vertex functions $\Phi$ in generally depend on the momenta, the magnetoelectrical couplings and microscopic data which parametrizes the classical order in the spin-wave theory, i.e. the angles $\phi$ and $\theta$. Due to the lengthy nature of the expressions we do not list them here explicitly, but they are straightforwardly obtained by using the Holstein-Primakoff expansion on $\mathcal{H}_\mathrm{E}$ (in the local frame).
It is however useful to note that (independtly of an explicit expression) the vertex functions need to obey the symmetrization and hermicity relations}
\begin{align}
  \Phi^{2,\bvec A,\mu}_{\alpha_1,\alpha_2}(\bm{k}) &=
    \Phi^{2,\bvec A,\mu}_{\alpha_2,\alpha_1}(\bvec{A}-\bm{k}) \\
  \Phi^{3,\bvec A,\mu}_{\alpha_1,\alpha_2,\alpha_3}(\bm{k}_1,\bm{k}_2,\bm{k}_3) &= (\bvec k_1,\alpha_1)\leftrightarrow(\bvec k_2,\alpha_2)\leftrightarrow(\bvec k_3,\alpha_3) \\
  \Phi^{2,\bvec A,\mu}_{\alpha_1,\alpha_2}(\bm{k})^\ast &=
    \Phi^{2,\bvec A,\mu}_{-\alpha_2,-\alpha_1}(\bm{k}-\bm{A}) \\
    \Phi^{3,\bvec A,\mu}_{\alpha_1,\alpha_2,\alpha_3}(\bm{k}_1,\bm k_2, \bm k_3)^\ast &=  \Phi^{3,\bvec A,\mu}_{-\alpha_1,-\alpha_2,-\alpha_3}(-\bm{k}_1,-\bm k_2,-\bm k_3),
\end{align}
for $\bvec A = 0,\bvec Q$.
\red{The explicit expansion in the local frame further yields that $\Phi^{2,0,\mu} = 0$.}
We diagonalize the quadratic spin-wave Hamiltonian $\heq^{(2)}$ by a Bogoliubov transformation, which transforms the spinor $\psi$ as
\begin{equation}
  \psi_{\alpha,\bvec k} = \Lambda_{\alpha \beta}(\bvec k)
  \hat{\psi}_{\beta,\bvec k} \label{eq:2}
\end{equation}
with the Bogoliubov matrix $\Lambda_{\alpha \beta} (\bvec k)$ given by
\begin{equation}
    \Lambda_{\bm{k}} = \begin{pmatrix} \cosh \vartheta_{\bm{k}} & \sinh
    \vartheta_{\bm{k}} \\ \sinh
    \vartheta_{\bm{k}} & \cosh \vartheta_{\bm{k}} \end{pmatrix}.
\end{equation}
The hyperbolic angle $\vartheta_\bvec k$ is determined by requiring that the anomalous terms in $\heq^{(2)}$ vanish, yielding $\tanh 2 \vartheta_{\bm{k}} = -(1-\delta/2) \gamma_{\bm{k}} / (2-\delta \gamma_{\bm{k}}/2)$.

We now include $\mathcal{H}_\mathrm{E}$ in the Keldysh action, giving a contribution of the form
\begin{equation}
  \mathcal{S}_\mathrm{E} = -\int\! \du\mathrm{t} \mathcal{H}_\mathrm{E}[E(t),\psi_+(t)] - \mathcal{H}_\mathrm{E}[E(t),\psi_-(t)],
\end{equation}
where $\psi_{\pm}(t)$ denote the fields on the forward and backward contours, respectively.
Defining the classical and quantum fields $\psi^{\cl,\qu} = (\psi_+ \pm \psi_-)/\sqrt{2}$, the Keldysh action for the interaction with the electric field reads
\begin{widetext}
\begin{align}
  \label{eq:SEfull}
  \mathcal{S}_E  = & -\int\! dt\, \sum_\mu E_\mu(t) \sum_{A=0,Q} \Bigg\{
        2 \sum_{\bm{k}} \hat\Phi^{2,A,\mu}_{\alpha_1,\alpha_2} (\bm{k})
        \hat\psi^\cl_{\alpha_1,\bm{k}} \hat\psi^\qu_{\alpha_2,A-\bm{k}}
        \nonumber\\
&    +  \frac{1}{\sqrt{2N}} \sum_{\bm{k}_1,\bm{k}_2,\bm{k}_3} \delta_{\sum_i \bm{k}_i,\bvec A} \,\hat\Phi^{3,\bvec A,\mu}_{\alpha_1,\alpha_2,\alpha_3}(\bm{k}_1,\bm{k}_2,\bm{k}_3)
          \left(
        \hat\psi^\qu_{\alpha_1,\bm{k}_1}\hat\psi^\qu_{\alpha_2,\bm{k}_2}\hat\psi^\qu_{\alpha_3,\bm{k}_3}+
        3
        \hat\psi^\qu_{\alpha_1,\bm{k}_1}\hat\psi^\cl_{\alpha_2,\bm{k}_2}\hat\psi^\cl_{\alpha_3,\bm{k}_3}\right) \Bigg\}.
\end{align}
\end{widetext}
\red{The effective action for the low-energy modes can now be obtained by introducing a energy cutoff and splitting the fields into slow and fast variables with respect to the cutoff, $\psi = \psi_< + \psi_>$. Expanding the weight $\eu^{\iu \mathcal S_\mathrm E}$ in the path integral and re-exponentiating yields to quadratic order 
\begin{equation} \label{eq:seff_deriv}
  \mathcal{S}_\mathrm{eff} = \langle \mathcal{S}_\mathrm{E} \rangle_> + \frac{\iu}{2} \left(\langle \mathcal{S}_\mathrm{E}^2 \rangle_> - \langle \mathcal{S}_\mathrm{E} \rangle^2_> \right) +  \dots,
\end{equation}
where $\langle \mathcal{O} \rangle_> = \int\!\mathcal{D}[\psi_>] \, \mathcal{O} \, \eu^{\iu \mathcal{S}} / \mathcal{Z}_>$ denotes functional averaging with respect to the fast modes of the system, i.e. modes above the cutoff, with $\mathcal{Z}_>$ being the appropriate partition function.
In practice, it is more convenient to replace the energy cutoff with a momentum cutoff $\lambda$.}
As noted in the main text, the lowest order non-trivial contribution to the effective action in second order perturbation theory is given by
\begin{equation}
   \mathcal{S}_\mathrm{eff} 
  =
  \begin{tikzpicture}[scale=2,baseline={([yshift=-.5ex]current bounding box.center)}]
    \path (1,0) coordinate (c);
    \path (c) ++(0:0.3) coordinate (a1); 
    \path (c) ++(0:0.6) coordinate (a2);
    \path (c) ++(150:0.6) coordinate (b); 
    \path (c) ++(180:0.3) coordinate (l1);
    \path (c) ++(210:0.6) coordinate (l2);
    \draw (c) circle (0.3);
    \draw[dashed] (a1)--(a2);
    \draw[dashed] (l1)--(b);
    \draw (l1)--(l2) node[left] {$\qu$};
    \draw (c) ++(30:0.33) node[above] {$\cl$};
    \draw (c) ++(150:0.33) node[above] {$\cl$};
    \draw (c) ++(-150:0.33) node[below] {$\cl$};
    \draw (c) ++(-30:0.33) node[below] {$\qu$};
  \end{tikzpicture},
\end{equation}
where have fixed the classical and quantum labels using that $G^{\qu\qu} = 0$ and the fact that the external leg must be quantum for the effective action to influence the low-energy dynamics of the system.

\red{Since the external leg is per definition a low-energy mode, it is convenient to work in the Holstein-Primakoff basis for the slow field, i.e. $\hat{\psi}^\qu_{\alpha_1} \rightarrow \psi^\qu_{\alpha_1}$, as the field $\psi$ can be directly related to the low-energy variables $m$ and $u$.
Accordingly, we then work with $\tilde{\Phi}^3_{\alpha_1, \alpha_2, \alpha_3} = \Lambda_{\alpha_2,\beta_1} \Lambda_{\alpha_3, \beta_2} \Phi^3_{\alpha_1, \beta_1,\beta_2}$ instead of $\hat{\Phi}^3$.
The internal lines in the diagram then correspond to contractions of $\hat{\psi}$, yielding Keldysh Green's functions for the magnons.
\begin{align}
  &\mathcal{G}_{\alpha\beta,\bm{k}}^{ab}(t-t')=
  -i \left\langle \hat\psi_{\alpha,\bm{k}}^{a}(t)
  \hat\psi_{\beta,\bm{k}'}^{b}(t')\right\rangle \delta_{\bvec k,-\bvec k'} \nonumber \\
  &\quad = \delta_{\bvec k,-\bvec k'} \left(\delta_{\alpha,1}\delta_{\beta,2} G_{\bm{k}}^{ab}(t-t') + \delta_{\alpha,2}\delta_{\beta,1} G^{ba}_{-\bm{k}}(t'-t)\right),
\end{align}
where $a,b=\mathrm{c},\mathrm{q}$ and the $G^{ab}$ are given by harmonic oscillator Green's functions 
\begin{align}
  \label{eq:75}
  G^{\cl\cl}_{\bm{k}}(t) & = G^K_{\bm{k}}(t) = -\iu \left(
                       2n_\mathrm{B}(E_{\bm{k}})+1\right) \eu^{-\iu
                       E_{\bm{k}} t}, \nonumber\\
  G^{\cl\qu}_{\bm{k}}(t) & = G^R_{\bm{k}}(t) = -\iu \Theta(t) \eu^{-\iu
                       E_{\bm{k}} t}, \nonumber\\
   G^{\qu\cl}_{\bm{k}}(t) & = G^A_{\bm{k}}(t) = \iu \Theta(-t) \eu^{-\iu
                        E_{\bm{k}} t}, \nonumber \\
  G^{\qu\qu}_{\bm{k}}(t) & = 0,
\end{align}
since $\hat{\psi} = (c_{\bvec k},c_{-\bvec k}^\dagger)^T$ involves the eigenmodes of $\mathcal{H}_\mathrm{eq}^{(2)}$. 
}

\subsection{Evaluation of loop diagram}

Up to a global factor, above diagram evaluates to
\begin{align} \label{eq:seff_fromloop}
  \mathcal{S}_\mathrm{eff}  =  -\iu \int \! \du t\, \du t' \, \sum_{\mu\nu} E_\mu(t) E_\nu(t') 
             \frac{1}{\sqrt{N}} \psi_{\alpha_1,\bm{0}}^q (t) \nonumber\\ 
      \times\sum_{\bm{k}} \tilde\Phi^{3,\bvec Q,\mu}_{\alpha_1,\alpha_2,\alpha_3}(0,\bm{k},\bm{Q}-\bm{k}) \hat\Phi^{2,\bvec Q,\nu}_{\beta_1,\beta_2}(-\bm{k}) \nonumber\\
      \times \mathcal{G}^{\cl\cl}_{\alpha_2\beta_1,\bm{k}}(t-t')\mathcal{G}^{\cl\qu}_{\alpha_3\beta_2,\bm{Q}-\bm{k}}(t-t').
\end{align}
In order to perform a $t$-integral in \eqref{eq:seff_fromloop}, it is convenient rewrite the electric field in terms of a complex amplitude $\mathcal{E}_\mu$ with
\begin{align}
  \label{eq:E_complex}
  E_\mu(t) & = \frac{1}{2} \left[\mathcal{E}_\mu \eu^{\iu\Omega t} +
             \mathcal{E}^*_\mu \eu^{-\iu\Omega t}\right].
\end{align}
Using the Fourier-transformed Green's functions $\mathcal{G}(t) = (2 \pi)^{-1} \int\! \du \omega \, \eu^{\iu \omega t} \mathcal{G}(\omega)$, we Fourier-transform \eqref{eq:seff_fromloop} to obtain
\begin{widetext}
\begin{align}
  \mathcal{S}_\mathrm{eff} = \frac{1}{4 \iu} \int\!\frac{\du \omega}{2 \pi} \frac{\du^2 \bvec k}{(2 \pi)^2} \, \tilde\Phi^{3,\bvec Q,\mu}_{\alpha_1,\alpha_2,\alpha_3} \hat\Phi^{2,\bvec Q,\nu}_{\beta_1,\beta_2} \bigg\{ &\Big[\psi^\qu_\alpha(0,0) \, \mathcal{E}_\mu \mathcal{E}_\nu^\ast \mathcal{G}^{\cl \cl}_{\alpha_2 \beta_1}(\bvec k,\omega) \mathcal{G}^{\cl \qu}_{\alpha_3 \beta_2}(\bvec Q-\bvec k,\Omega-\omega) + (\Omega \to -\Omega, \mathcal{E}_\mu \leftrightarrow \mathcal{E}_\nu) \Big] \nonumber \\
  + &\Big[\psi^\qu_\alpha(0,2 \Omega) \, \mathcal{E}_\mu \mathcal{E}_\nu \mathcal{G}^{\cl \cl}_{\alpha_2 \beta_1}(\bvec k,\omega) \mathcal{G}^{\cl \qu}_{\alpha_3 \beta_2}(\bvec Q-\bvec k,-\Omega-\omega)  + (\Omega \to -\Omega, \mathcal{E} \rightarrow \mathcal{E}^\ast)\Big]\bigg\},
\end{align}
\end{widetext}
where the vertex functions have the same momentum dependence as in \eqref{eq:seff_fromloop}.
Clearly, the external quantum field in the terms in the second bracket has frequency $2 \Omega$ which is above any cutoff frequency $\omega_0$ and therefore does not contribute to the dynamics of the slow variables $u$ $m$.
We therefore drop these terms and proceed with the $\omega$ integration by exploiting the fact that $G^K_\bvec k(\omega) = - 2 \iu \pi (2 n_\mathrm{B} (E_\bvec k) + 1) \delta(\omega- E_\bvec k)$.
The four terms arising from non-vanishing possible combinations of the particle-hole indices $\alpha_i$ and $\beta_i$ can be conveniently expressed by letting the indices take values $1\rightarrow +$, $2\rightarrow -$.
The result can be cast into the form
\begin{equation}
  \mathcal{S}_\mathrm{eff} = \int\! \du t\, \du^2 \bvec x \, h_\alpha \psi_\alpha^\qu(\bvec x, t),
\end{equation}
which defines the spatially homogenous effective field $h_\alpha$ and leads to Eq. \eqref{eq:h_full} in the main text.

\subsection{Analytical momentum integration}

The momentum integral in the expression for the induced field \eqref{eq:h_full} can be performed analytically for small momenta $\lambda \leq \bvec k < \Lambda$, where $\lambda$ is an IR cutoff which would correspond to a low-frequency cutoff and $\Lambda$ is a UV cutoff which is determined by the lattice spacing.

We also note that a small $\delta$ is sufficient to gap out the magnon dispersion at $\bvec Q$, allowing to use $\delta \ll 1$ as a perturbative parameter.
The momentum space integrals are then conveniently done in polar coordinates $\bvec k = (k,\alpha)$, with $\du^2 \bvec k = k \, \du k \, \du \alpha$.
Expanding the magnon dispersion in \eqref{eq:magnon_disp} in $\delta$ and then to linear order in $\bvec k$, we find that
\begin{subequations} \label{eq:disp_expand}
  \begin{align}
  \epsilon_{\bvec k} &\simeq 2 \sqrt{2} J k - J k \delta /\sqrt{2} \\
  \epsilon_{\bvec Q - \bvec k} &\simeq \frac{4 \sqrt{2} \delta  J}{k}+\left(2 \sqrt{2} J + \frac{\delta  J \cos (4 \alpha )}{12 \sqrt{2}}-\frac{9 \delta  J }{4 \sqrt{2}}\right)k.
\end{align}
\end{subequations}
Note that the magnon dispersion at $\bm Q$ becomes singular at finite $\delta$ since the dispersion is quadratic at $\delta > 0$. However potential divergences may be regularized by a cutoff $\lambda > 0$.

We proceed similarly for the vertex functions in the nominator of \eqref{eq:h_full}.
The product of $\hat{\Phi}^{2,\bvec Q,v}$ and $\tilde{\Phi}^{3,\bvec Q,v}$ will in general contain four Bogoliobuv factors, two of which depend on $\Theta_{\bm{k}}$ and two on $\Theta_{\bm{Q}-\bm k}$, respectively.
The small $\delta$ and low-energy expansion can be performed conveniently after rewriting the product as a function of $\sinh 2 \Theta$ and $\cosh 2 \Theta$.

After using the Dirac identity $(x+\iu \eta)^{-1} = - \iu \pi \delta(x) + \pv 1/x$ to split the fields into resonant and non-resonant parts, the integration for the resonant parts can be performed analytically by using that
\begin{equation} \label{eq:delta_integral}
  \delta\left(\Omega - \epsilon_{\bm k} - \epsilon_{\bm{Q}-\bm{k}}\right) = \sum_{i} \frac{\delta(k-k_i)}{\partial_k (\epsilon_{\bm k}+\epsilon_{\bm Q - \bm k})|_{k_i}},
\end{equation}
where $k_i = k_i(\alpha)$ are the roots of $\Omega = \epsilon_\bvec k + \epsilon_{\bvec Q - \bvec k}$, given by
\begin{align}
  k_1(\alpha) &= \frac{4 \sqrt{2} \delta  J}{\Omega }, \\
  k_2(\alpha) &= \frac{\Omega }{4 \sqrt{2} J}-\frac{  \Omega ^2 \cos 4 \alpha +3072 J^2-39 \Omega ^2}{384 \sqrt{2} J \Omega } \delta.
\end{align}
Note that, since we are working with $\delta \ll 1$, $k_1$ is expected to be small and thus lie below the IR cutoff $\lambda$. We therefore proceed with $k_2(\alpha)$.
Using \eqref{eq:delta_integral}, the $k$-integration is trivial, and the angular integrals are elementary.
For the non-resonant part, it is convenient to perform the angular integration first and then the $k$-integration.
While the integrals are logarithmically divergent, it is safe to take the cutoff $\lambda \to 0$, as $\Omega$ regularizes the IR divergence, however we find it necessary to keep a finite UV cutoff $\Lambda$.

\subsection{Analytical expressions for effective fields} \label{sec:analy_fields}

The results to linear order in $\delta$ read
\begin{widetext}
\begin{align}
  h_m &= h_m^\mathrm{R} + h_m^\mathrm{NR} \quad \text{with} \\
  h_m^\mathrm{R} &= \bigg( g_1 \mathcal{N}_{(0,0)} \cos (2 \phi ) \left(\mathcal{E}_y \bar{\mathcal{E}}_x+\mathcal{E}_x \bar{\mathcal{E}}_y\right) 
  +\frac{1}{2} \left(\mathcal{E}_x \bar{\mathcal{E}}_x - \mathcal{E}_y \bar{\mathcal{E}}_y \right) \left(g_2-g_3\right) \mathcal{N}_{(0,0)} \sin (2 \phi ) \bigg) \left(-2 g_5 \sin (2 \theta )+\left(g_2+g_3\right) \cos (2 \theta )+2 g_4\right) \nonumber\\
  &+\frac{1}{4} \left(\mathcal{E}_x \bar{\mathcal{E}}_x + \mathcal{E}_y \bar{\mathcal{E}}_y \right) \left(4 g_1^2-\left(g_2-g_3\right){}^2\right) \mathcal{N}_{(0,0)} \sin (4 \phi ) \nonumber\\
   &+ \delta \biggl[ \bigg( -g_1 \cos (2 \phi) \left(\mathcal{E}_y \bar{\mathcal{E}}_x+\mathcal{E}_x \bar{\mathcal{E}}_y\right) 
   -\frac{1}{2} \left(\mathcal{E}_x \bar{\mathcal{E}}_x - \mathcal{E}_y \bar{\mathcal{E}}_y \right) \left(g_2-g_3\right) \sin (2 \phi) \bigg) \nonumber\\
   &\qquad \times\left(-2 g_5 \mathcal{N}_{(1,0)} \sin (2 \theta )+ (g_2 + g_3) \mathcal{N}_{(1,0)} \cos (2 \theta )+g_4 \mathcal{N}_{(1,1)}\right) -\frac{1}{4} \left(\mathcal{E}_x \bar{\mathcal{E}}_x + \mathcal{E}_y \bar{\mathcal{E}}_y \right) \left(4 g_1^2-\left(g_2-g_3\right){}^2\right) \mathcal{N}_{(1,0)} \sin (4 \phi ) 
   \biggl]\\
  h_m^\mathrm{NR} &= \frac{g_1 \left(g_2-g_3\right)}{3\times 2^{21} \pi J^5} \iu\left(\mathcal{E}_x \mathcal{\bar{E}}_y-\mathcal{E}_y \mathcal{\bar{E}}_x\right) \bigg(  \left(1024 J^2 \Omega ^2+3 \Omega ^4\right) \mathcal{L}_{(0)} \nonumber\\
     &\qquad+ \left(768 J^4 \Lambda ^4+16384 J^4 \Lambda ^2+128 \sqrt{2} J^3 \Omega  \Lambda ^3+4096 \sqrt{2} J^3 \Omega  \Lambda +48 J^2 \Omega ^2 \Lambda ^2+12 \sqrt{2} J \Omega ^3 \Lambda \right)\bigg)\nonumber\\
     &+ \delta \frac{g_1 \left(g_2-g_3\right)}{3^4 \times 2^{27} \pi  J^5 \left(8 J \Lambda -\sqrt{2} \Omega \right)} \iu\left(\mathcal{E}_x \bar{\mathcal{E}}_y-\mathcal{E}_y \bar{\mathcal{E}}_x\right) \biggl[ \big(4306944 J^5 \Lambda ^5+70778880 J^5 \Lambda ^3 +452984832 J^5 \Lambda \nonumber\\
     &\quad+897280 \sqrt{2} J^4 \Omega  \Lambda ^4+26542080 \sqrt{2} J^4 \Omega  \Lambda ^2+448640 J^3 \Omega ^2 \Lambda ^3-13271040 J^3 \Omega ^2 \Lambda +168240 \sqrt{2} J^2 \Omega ^3 \Lambda ^2-84120 J \Omega ^4 \Lambda \big) \nonumber\\
     &\quad +\biggl(-3^3 \times 2^{24} J^5 \Lambda +3^3 \times 2^{21} \sqrt{2} J^4 \Omega +13271040 J^3 \Omega ^2 \Lambda \nonumber\\
     &\qquad -1658880 \sqrt{2} J^2 \Omega ^3+84120 J \Omega ^4 \Lambda -10515 \sqrt{2} \Omega ^5\biggl) \mathcal{L}_{(0)}\biggl]
\end{align}
and
\begin{align}
  h_u &= h_u^\mathrm{R} + h_u^\mathrm{NR} \nonumber\\
  h_u^\mathrm{R} &= \left\{\frac{ \Omega ^3 (1+\delta) }{3 \times 2^{15} \sqrt{2} J^4}  \right\}g_1 (g_2-g_3)\,\iu\left(\mathcal{E}_x \bar{\mathcal E}_y-\mathcal E_y \bar{\mathcal E}_x\right) \\
  h_u^\mathrm{NR} &= \bigg( 2 g_1 \mathcal{M}_{(0,0)} \cos (2 \phi ) \left(\mathcal{E}_y \bar{\mathcal{E}}_x+\mathcal{E}_x \bar{\mathcal{E}}_y\right) 
  +\left( \mathcal{E}_x \bar{\mathcal{E}}_x - \mathcal{E}_y \bar{\mathcal{E}}_y \right) \left(g_2-g_3\right) \mathcal{M}_{(0,0)} \sin (2 \phi ) \bigg) \left(-2 g_5 \sin (2 \theta )+\left(g_2+g_3\right) \cos (2 \theta )+2 g_4\right)\nonumber\\
  &\qquad+\frac{1}{2} \left( \mathcal{E}_x \bar{\mathcal{E}}_x + \mathcal{E}_y \bar{\mathcal{E}}_y \right) \left(4 g_1^2-\left(g_2-g_3\right){}^2\right) \mathcal{M}_{(0,0)} \sin (4 \phi )
  \nonumber\\
  &+\delta \biggl[\bigg( g_1 \cos (2 \phi ) \left(\mathcal{E}_y \bar{\mathcal{E}}_x+\mathcal{E}_x \bar{\mathcal{\mathcal{E}}}_y\right) 
  +\frac{1}{2} \left( \mathcal{E}_x \bar{\mathcal{E}}_x - \mathcal{E}_y \bar{\mathcal{E}}_y \right) \left(g_2-g_3\right) \sin (2 \phi ) \bigg) \nonumber\\ 
  &\qquad \times \left(-8 g_5 \mathcal{M}_{(1,0)} \sin (2 \theta )+4 \left(g_2+g_3\right) \mathcal{M}_{(1,0)} \cos (2 \theta )+\sqrt{2} g_4 \mathcal{M}_{(1,1)}\right) \nonumber\\
  &\qquad+\left( \mathcal{E}_x \bar{\mathcal{E}}_x + \mathcal{E}_y \bar{\mathcal{E}}_y \right) \left(4 g_1^2-\left(g_2-g_3\right){}^2\right) \mathcal{M}_{(1,0)} \sin (4 \phi ) \biggl]
  \end{align}
where we have defined the common prefactors
\begin{align}
  \mathcal{N}_{(0,0)} &= \frac{\Omega ^2 \left(1024 J^2+3 \Omega ^2\right)}{3 \times 2^{21} J^5} \\
  \mathcal{N}_{(1,0)} &= \frac{9 \times 2^{21} J^4-2 \times 3^3 \times 5 \times 2^{15} J^2 \Omega ^2 - 3505 \Omega ^4}{3^3 \times 2^{27} J^5} \\
      \mathcal{N}_{(1,1)} &= \frac{3^3 \times 2^{21} J^4-5 \times 3^3 \times 2^{12} J^2 \Omega ^2 - 3505 \Omega^4}{3^3 \times 2^{26} J^5} \\
    \mathcal{L}_{(0)} &= \log \left|\frac{\Omega -4 \sqrt{2} J \Lambda}{\Omega}\right| \\
    \mathcal{M}_{(0,0)} &= \frac{1}{294912 \pi  J^4} \left[-128 \sqrt{2} J^3 \Lambda^3-48 J^2 \Omega  \Lambda ^2-3 \Omega ^3 \mathcal{L}_{(0)} -12 \sqrt{2} J \Omega ^2 \Lambda \right] \\
    \mathcal{M}_{(1,0)} &= \frac{-224 J^4 \Lambda ^4+64 \sqrt{2} J^3 \Omega  \Lambda ^3+48 J^2 \left(64 J^2+\Omega ^2\right) \Lambda ^2-12 \sqrt{2} J (\mathcal{L}_{(0)}+1) \Omega ^3 \Lambda +3 \mathcal{L}_{(0)} \Omega ^4}{589824 \pi J^4  (\Omega- 4 \sqrt{2} J \Lambda)} \\
    \mathcal{M}_{(1,1)} &= \frac{3328 \sqrt{2} J^4 \Lambda ^4+1664 J^3 \Omega  \Lambda ^3+ \left(512 J^2-13 \Omega ^2\right) \left( 24 J (\mathcal{L}_{(0)}+1) \Omega   \Lambda -3 \sqrt{2} \mathcal{L}_{(0)} \Omega ^2 -48 \sqrt{2} J^2  \right) \Lambda ^2}{9 \times 2^{17} \pi J^4 (\Omega- 4 \sqrt{2} J \Lambda)}
\end{align} 
\end{widetext}

\section{Driving with effective fields} \label{sec:intpump}

To inspect how the effective fields act on the dynamics of the low-energy equations of motion, we focus on the equation of motion for the magnetization $m(t)$, which is given by
\begin{equation} \label{eq:EOM_ODE}
  \partial^2_t m + 2 \gamma \partial_t m + \omega_0^2 m = \kappa h_m + \partial_t h_u,
\end{equation}
corresponding to a driven harmonic oscillator with $\gamma = 1/ 2 \tau$ and $\omega_0^2 = \kappa / \chi$, and we define $\kappa = 8 \Gamma S$.
For the homogenous solution we impose the initial conditions $m(0) = \partial_t m(0) = 0$, yielding the trivial solution $m=0$.
We now consider the inhomogeneous source terms. Since \eqref{eq:EOM_ODE} is linear, the two contributions can be considered separately.
Using the Green's function of the harmonic oscillator
\begin{equation}
  G(t-t') = -\int_{-\infty}^\infty \! \frac{\mathrm{d}\omega}{2 \pi} \frac{\eu^{-\iu \omega (t-t')}}{\omega^2+ 2 i \gamma \omega - \omega_0^2}
\end{equation}
the motion of the magnetziaton $m_m(t)$ and $m_u(t)$ in the presence of a (time-dependent) external field $h_m$ and $h_u$, respectively, is obtained conveniently by a convolution
\begin{align} \label{eq:convolution}
  m_m(t) &= \int_{-\infty}^\infty \! \du t' \, h_m(t') G(t-t') , \\
  m_u(t) &= -\int_{-\infty}^\infty \! \du t' \,  h_u(t') \partial_{t'} G(t-t'),
\end{align}
where we have partially integrated the inhomogeneity in the equation of motion for $m_u$, using that $h_u(\pm \infty) = 0$.
We emphasize that this approach straightforwardly yields the particular solutions $m_{m,u}(t)$ for all times $t$. The damping term imposes causality.
We consider unit pulses of strength $\bar{h}$ starting at $t=0$ of duration $t_p$
\begin{equation}
  h_{m,u}(t) = \bar{h}_{m,u} \left(\Theta(t) - \Theta(t-t_p) \right).
\end{equation}
After some algebra, \eqref{eq:convolution} yields 
\begin{widetext}
\begin{align}
  m_m(t) &=\kappa \bar{h}_m \left[\Theta (t) 
  \frac{\bar{\omega} - \eu^{-\gamma  t} (\gamma  \sin \bar{\omega} t+\bar{\omega}  \cos \bar{\omega} t)}{\bar{\omega}  \left(\gamma ^2+\bar{\omega} ^2\right)}-
  \Theta (t-t_p ) 
  \frac{\bar{\omega} - \eu^{-\gamma  (t-t_p )} (\gamma  \sin \bar{\omega}  (t-t_p )+\bar{\omega}  \cos \bar{\omega}  (t-t_p ))}{\bar{\omega}  \left(\gamma ^2+\bar{\omega} ^2\right)}\right] \\
  m_u(t) &= \bar{h}_u \Theta (t) \frac{ \eu^{-\gamma t}  \sin \bar{\omega} t}{\bar{\omega} }-\bar{h}_u \Theta (t-t_p )  \frac{\eu^{-\gamma  (t-t_p )} \sin \bar{\omega}  (t-t_p )}{\bar{\omega} },
\end{align}
\end{widetext}
where $\bar{\omega} = \sqrt{\omega_0^2 - \gamma^2}$ is the eigenfrequency of the system with damping.
After time $t=t_p$, when the pulse is turned off, $m(t)$ can be seen to evolve according to the homogenous equations of motion, with the initial conditions obtained by requiring continuity at time $t_p$.
The initial conditions can be obtained by evaluating $m_{m,u}(t)$  as well as their time derivatives shortly after the pulse, i.e. $t= t_p^+ = t_p + 0^+$.
We find that
\begin{align}
  m_m(t_p^+) &= \kappa \bar{h}_m  \frac{\eu^{-\gamma  t_p } \left(\bar{\omega}  \eu^{\gamma  t_p }-\gamma  \sin (t_p  \bar{\omega} )-\bar{\omega}  \cos (t_p  \bar{\omega} )\right)}{\bar{\omega}  \left(\gamma ^2+\bar{\omega} ^2\right)}\nonumber \\
  &= \mathcal{O}(t_p^2) \label{eq:initm0} \\
  \partial_t m_m(t_p^+) &=  \kappa \bar{h}_m \frac{\eu^{-\gamma  t_p } \sin   \bar{\omega} t_p}{\bar{\omega} } =  \bar{h}_m \kappa t_p + \mathcal{O}(t_p^2) \label{eq:initm1},
\end{align}
where we have expanded the resulting expressions for the initial amplitude and velocity for short pulses $t_p \ll \bar{\omega}^{-1}, \gamma^{-1}$.
Proceeding equivalently for $h_u$ gives
\begin{align}
  m_u(t_p^+) &=\bar{h}_u  \frac{\eu^{-\gamma  t_p } \sin \bar{\omega} t_p}{\bar{\omega} } = \bar{h}_u t_p + \mathcal{O}(t_p^2) \label{eq:initu0}\\ 
  \partial_t m_u(t_p^+) &= -\bar{h}_u\frac{\eu^{-\gamma  t_p }  \left(\bar{\omega}  \eu^{\gamma  t_p }+\gamma  \sin   \bar{\omega} t_p-\bar{\omega}  \cos \bar{\omega} t_p\right)}{\bar{\omega} } \nonumber \\ &= \mathcal{O}(t_p^2) \label{eq:initu1}
\end{align}
We thus find that the effective field $h_m$ in the ultrafast regime acts as an impulse to the system, giving the magnetization an initial velocity $\partial_t m$, while the field $h_u$ displaces the magnetization and thus provides an initial amplitude for the free oscillations.

In addition to the ultrafast regime $t_p \ll \bar{\omega}^{-1},\gamma^{-1}$, we may in addition consider the regime in which the pulse is short compared to the relaxation timescale $t_p \ll \gamma^{-1}$, but on the order of the magnon oscillation time period (note that $\bar{\omega} > \gamma$ for oscillations to occur). The initial conditions in this regime are straightforwardly obtained from Eqs. \eqref{eq:initm0}-\eqref{eq:initu1} by letting $\eu^{\pm\gamma t_p} \to 1$.

Third, we consider long pulses $t_p \gg \bar{\omega}^{-1},t_p^{-1}$. In this regime, one obtains
\begin{align}
  m_m(t_p^+) &\to \frac{\kappa \bar h_m}{\gamma^2 + \bar{\omega}^2}, \quad \partial_t m_m(t_p^+) \to 0 \\
  m_u(t_p^+) &\to 0, \quad \partial_t m_u(t_p^+) = -\bar h_u.
\end{align}
Therefore, in the limit of long pulses, the role of the effective fields is  reversed: $h_m$ sets the initial amplitude. This corresponds to the magnetization relaxing to a non-zero equilibrium value \emph{during} the pump due to the constant force $h_m$ acting.
Conversely, $h_u$ acts an impulse which gives an initial velocity to the oscillations -- this can be seen from the fact that $h_u$ enters the EOM via its time derivative $\partial_t h_u$, so that the discontinuous switching off acts as a $\delta$-kick, providing an initial velocity.

\section{Magnetoelectrical couplings} \label{sec:MEcoup}

In this appendix, in order to make the presentation self-contained, we summarize the method of microscopic calculation due to Bolens,\cite{bolens} which we used to estimate parameters. For more details please see Ref.~\onlinecite{bolens}.

The couplings in $g_1 \dots g_5$ in \eqref{eq:ham_E} can be related to magnetoelectrical couplings through the definition of a spin-dependent polarization $\vec P = \vec P(\{{\sf S}\})$, so that $\mathcal{H}_\mathrm{E}$ can be written
\begin{equation}
  \mathcal{H}_\mathrm E = -\sum_i \vec P \cdot \vec E.
\end{equation}
Note that in the present geometry, we assume the light to propagate normal to the IrO planes, so that $\vec E = (E_x, E_y, 0)^T$.
In a microscopic treatment, matrix elements of the quantum-mechanical polarization operator can be evaluated in a tight-binding approach, in which the respective Wannier functions are approximated by individual atomic orbitals.
Matrix elements of $\vec P$ for ions at sites $\vec R_i$ and $\vec R_j$ are then given by
\begin{equation}
  \braket{\vec R_i ,A | \vec P | \vec R_j',B} =  \braket{\vec R_i,A | e \vec r | \vec R_j',B},
\end{equation}
where $r$ ist the position operator and $e$ the electric charge, and $A,B$ denoting any additional indices such as orbital or spin degrees of freedom. The spin-dependent polarization is defined in the low-energy subspace given by singly occupied sites, akin to the derivation of the superexchange interaction from the Hubbard model.

To this end, we use the approach by Bolens and briefly review the microscopic model introduced in Ref.~\onlinecite{bolens}, which is given by a three-band Hubbard model for the $t_{2g}$ manifold,
\begin{equation} \label{eq:full_hub_h}
  \mathcal{H} = \mathcal{H}_\mathrm{hop} + \mathcal{H}_\mathrm{CF} + \mathcal{H}_\mathrm{SOC} + \mathcal{H}_\mathrm{int},
\end{equation}
where the hopping (from both direct and indirect processes) between the $\mathrm{Ir}$ ions at site $i$ and $j$ and with orbitals $a,b\in yz(x),xz(y),xy(z)$ has the spin-diagonal ($\sigma=\uparrow,\downarrow$) form
\begin{equation}
  \mathcal{H}_\mathrm{hop} = - \sum_{\langle i j\rangle} c_{i,a,\sigma}^\dagger t_{i,a;j,b} c_{j,b,\sigma},
\end{equation}
and spin-orbit coupling of strength $\lambda$ and tetragonal crystal field splitting lead to two onsite terms, 
\begin{subequations}
  \begin{align}
    \mathcal{H}_{\mathrm{SOC}} &= \lambda\sum_{i,\alpha} c_{i,a,\sigma}^\dagger  L^\alpha_{a,b} S^\alpha_{\sigma,\sigma'} c_{i,b,\sigma'} \quad \text{and} \label{eq:h_CF} \\
    \mathcal{H}_{\mathrm{CF}} &= \Delta\sum_{i,\alpha} c_{i,a,\sigma}^\dagger  [(L^z)^2]_{a,b} \delta_{\sigma,\sigma'} c_{i,b,\sigma'}, \label{eq:h_SOC}
  \end{align}
\end{subequations}
where the $L=1$ matrices are given by $L^a = -\iu \epsilon_{abc}$ and $S^\alpha = \tau^\alpha/2$ with the Pauli matrices $\tau^\alpha$ as usual.
Note that the orbitals above are defined in a local basis, which due to the octahdral rotation are rotated at an angle $\alpha$ with the bond direction in the plane.   
Intra- and and interorbital Coulomb repulsion $U$ and $U' = U - 2 J_\mathrm{H}$ respectively, and the Hund's coupling $J_\mathrm H$ give rise to interacting terms in the Hamiltonian,
\begin{align}
  \mathcal{H}_\mathrm{int} = &\sum_{i,a} n_{i,a,\uparrow} n_{i,a,\downarrow} + (U' - J_\mathrm H) \sum_{i,a < b ,\sigma}  n_{i,a,\sigma} n_{i,b,\sigma} \nonumber\\
  &+ U'\sum_{i,a \neq b} n_{i,a,\uparrow} n_{i,b,\downarrow} - J_\mathrm{H} \sum_{i,a \neq b} c_{i,a,\uparrow}^\dagger c_{i,a,\downarrow} c_{i,b,\downarrow}^\dagger c_{i,b,\uparrow} \nonumber\\
  &+ J_\mathrm H \sum_{i,a \neq b} c_{i,a,\uparrow}^\dagger c_{i,a,\downarrow} c_{i,b,\downarrow} c_{i,b,\uparrow}.
\end{align}
We now consider an Ir-Ir $x$-bond, with the assumption that only indirect hopping (via the oxygen $p$-orbitals) occurs.
The hopping thus takes the form $[t_{ij}]_{a,b} = \diag(0,t_2,t_2 \cos 2\theta)_{a,b}$, with $t_2 = - t_{pd\pi}^2/\Delta_{pd}$, where $\Delta_{pd}$ is the charge-transfer energy, and $t_{pd\pi}$ is a Slater-Koster parameter.\cite{slakost}
The low-energy manifold is spanned by states with one hole per site, and $\lambda > 0$ and $\Delta > 0$ further split degeneracies so that there is a unique $J_\mathrm{eff} = 1/2$ Kramers doublet ground state per site, denoted with $\ket{\sigma_i}$ with $\sigma = \uparrow,\downarrow$ and obtained by diagonalizing $\mathcal{H}_\mathrm{SOC} + \mathcal{H}_\mathrm{CF}$ for a single site.
The $z$-component $J^z$ of the total angular momentum $\vec J = \vec L + \vec S$ commutes with $\mathcal{H}_{\mathrm{CF}+\mathrm{SOC}}$, and can therefore be used as a quantum number to label the two-fold degenerate ground states.
A  basis for the four-dimensional low-energy subspace (for the two Ir ions on the bond) is then given by $\ket{\sigma_1}\otimes\ket{\sigma_2}$ (see also the similar approach outlined in Ref.~\onlinecite{khaliu12}).
In a second quantized notation and after integrating out the $p$-orbitals, the components of the hopping polarization read
\begin{equation}
  P^\alpha = \sum_\alpha  c_{i,a,\sigma}^\dagger \left(\mathcal{P}^\alpha_S + \mathcal{P}^\alpha_A\right)_{a,b} c_{j,b,\sigma}
\end{equation}
with relevant $\alpha = x,y$ components of the matrices $\mathcal{N}^\alpha_{S/A}$ given by
\begin{subequations}
  \begin{align}
    \left[\mathcal{P}^x_S\right]_{a,b} &= \sin (\theta) t_{pd\pi} P^\perp_{p \pi d \sigma} |\epsilon_{zab}| \\
    \left[\mathcal{P}^x_A\right]_{a,b} &= 0 \\
    \left[\mathcal{P}^y_S\right]_{a,b} &= -\diag[0,P^\parallel_{pd\pi},P^\perp_{p \sigma d \pi} + (P^{\parallel}_{pd\pi} + P^{\perp}_{p\sigma d \pi}) \cos 2 \theta]_{a,b} \nonumber\\
    &\qquad \times 2 \sin(\theta) t_{pd\pi} \\
    \left[\mathcal{P}^y_A\right]_{a,b} &= - \cos (\theta) \, t_{pd\pi} P^\sigma_{p\pi d \sigma} \epsilon_{zab},
  \end{align}
\end{subequations}
where $P^\perp_{p \pi d \sigma}$ etc. denote the orbital polarization integrals in the Slater-Koster notation, which we approximate by hydrogenlike wavefunctions, which is only a crude approximation for the spread out $5d$ orbitals.
To correct for this, we maximize the orbital overlap by assuming zero interatomic distance,\cite{knb,bolens} yielding
\begin{equation}
  P_{pd\pi}^\parallel = P^\perp_{p \sigma d \pi}  =   0.1619 \quad \text{and} \quad P^\sigma_{p\pi d \sigma} = -0.0935.
\end{equation} 
A more sophisticated approach would require the evaluation of the polarization integrals for the relevant orbitals from first principles.
Denoting the matrix representation of the Hamiltonian on the full Hilbert space with $\uvec{\mathcal{H}}$, the matrix representation of the polarization in the pseudospin basis is then obtained as
\begin{equation}
  \uvec{P}_\mathrm{eff}^\alpha = \uvec{\mathbb P}^\dagger \uvec{U}^\dagger \uvec{P}^\alpha \uvec{U} \, \uvec{\mathbb{P}},
\end{equation}
where $\uvec{U}$ diagonalizes $\uvec{\mathcal{H}}$ and $\uvec{\mathbb P}$ is the matrix representation of the projection operator from the full eigenbasis to the basis of the low-energy subspace (note that the states after the projection need to be normalized).
Various coefficients $g_i$ in \eqref{eq:ham_E} are then obtained by computing the matrix scalar product $\mathrm{Tr}[\uvec{P}_\mathrm{eff}^\alpha \uvec{S}_1^\beta \uvec{S}_2^\gamma]$ for the respective $\beta,\gamma$, and transforming back to the global frame $S\to {\sf S}$.
Note that in general also a spin-independent contribution to $\vec P$ occurs, which we neglect for our purposes.

For the numerical evaluation, we employ the parameters of Ref.~\onlinecite{khaliu12}, $\Delta_{pd} = 3.3 \, \mathrm{eV}$, $\Delta_{xy} = 0.15 \, \mathrm{eV}$, $U= 1.86 \, \mathrm{eV}$, $\lambda = 0.4 \, \mathrm{eV}$, $t_{p d \pi} = 0.83 \, \mathrm{eV}$, and $\theta=11^\circ$ and thus obtain \eqref{eq:mecoup}.
Note that $g_1$ is suppressed since the relevant polarization integral $P^\perp_{p \pi d \sigma}$ is an order of magnitude smaller.
Furthermore, we find that in the limit of small $J_\mathrm H$ and small $\lambda$ that $g_2 = g_3$, which is consistent with analytical perturbative approaches in this limit.\cite{bolens}
As the inverse Faraday effect is proportional to $g_2-g_3$, we emphasize that a sizeable $J_\mathrm{H}$ and $\lambda$ are necessary for the effect.



\begin{thebibliography}{99}
\bibitem{fn:numerics}
Working at $T=0$, we have employed a momentum space discretization of $200 \times 200$ points and used a broadening $\eta = 0.03$ for numerical convenience, having checked that different $\eta$ and discretizations give comparable results.
\bibitem{fn:nonzero}
For clarity, we only show non-zero components.
\end{thebibliography}
\end{document}